\documentclass[submission, Phys]{SciPost}

\hypersetup{
    colorlinks,
    linkcolor={red!50!black},
    citecolor={blue!50!black},
    urlcolor={blue!80!black}
}

\usepackage[bitstream-charter]{mathdesign}
\usepackage{bm}
\usepackage{amsthm}

\newtheorem*{lemmanon}{Lemma}

\def\infinity{\infty}

\def\:={\,\raisebox{0.85pt}{.}\hspace{-2.78pt}\raisebox{2.85pt}{.}\!\!=\,}
\def\=:{\,=\!\!\raisebox{0.85pt}{.}\hspace{-2.78pt}\raisebox{2.85pt}{.}\,}

\DeclareSymbolFont{usualmathcal}{OMS}{cmsy}{m}{n}
\DeclareSymbolFontAlphabet{\mathcal}{usualmathcal}

\begin{document}

\begin{center}{\Large \textbf{
Quantum adiabaticity in many-body systems and almost-orthogonality in complementary subspace\\
}}\end{center}

\begin{center}
Jyong-Hao Chen \textsuperscript{1,2$\star$} 
and
Vadim Cheianov \textsuperscript{1}
\end{center}

\begin{center}
{\bf 1} Instituut-Lorentz, Universiteit Leiden, P.O.\ Box 9506, 2300 RA Leiden, The Netherlands
\\
{\bf 2} Department of Physics, National Central University, Chungli 32001, Taiwan
\\
${}^\star$ {\small \sf jyonghaochen@gmail.com}
\end{center}

\begin{center}
\today
\end{center}


\section*{Abstract}
{\bf
We investigate why, in quantum many-body systems, the adiabatic fidelity and the overlap between the initial state and instantaneous ground states 
often yield nearly identical values. 
Our analysis suggests that this phenomenon results from an interplay between two intrinsic limits of many-body systems: 
the limit of small evolution parameters and the limit of large system sizes. 
In the former case, conventional perturbation theory provides a straightforward explanation. 
In the latter case, a key insight is that pairs of vectors in the Hilbert space orthogonal to the initial state tend to become nearly orthogonal as the system size increases. 
We illustrate these general findings with two representative models of driven many-body systems: the driven Rice-Mele model and the driven interacting Kitaev chain model.
}

\vspace{10pt}
\noindent\rule{\textwidth}{1pt}
\tableofcontents\thispagestyle{fancy}
\noindent\rule{\textwidth}{1pt}
\vspace{10pt}

\section{Introduction}
\label{sec: introduction}

In the modern era of quantum technologies, it is crucial to prepare and manipulate quantum states with precision. 
For this purpose, various approximations based on the {\it quantum adiabatic theorem} (QAT) \cite{Born27,Born28,Kato50,Messiah14} are widely used.
Examples of applications include adiabatic quantum transport \cite{Thouless83,Berry84,Avron88,Avron95}, 
adiabatic quantum computation \cite{Farhi00,Roland02,Albash18}, 
and adiabatic quantum state manipulation \cite{Ivanov01,Nayak08,Hamma08,Alicea11,Halperin12,Heck12}
and preparation \cite{Guzik05,Aharonov07s,Wecker15,Reiher17,Wan20,Perez22}.
Adiabatic evolution refers to the evolution of a quantum system whose time-evolved state remains close to its instantaneous eigenstate.
A well-known adiabatic criterion \cite{Messiah14,Jansen07} 
\footnote{
See also Ref.\ \cite{Albash18} for a comprehensive review.}
is that the rate of change of the time-dependent Hamiltonian 
must be much smaller than certain non-negative powers of the minimum energy gap of the Hamiltonian.
Although calculating energy gaps is often possible for systems with a small Hilbert space or for exactly solvable models, 
it is generically difficult for many-body systems with a large Hilbert space.
Alternatively, the formalism of {\it shortcuts to adiabaticity} (STA) promises to give the same adiabatic results as those provided 
by the QAT but without requiring slow driving \cite{Demirplak03,Berry09,Chen10,Torrontegui13,Jarzynski13,Campo13,Bukov19,Odelin19}.
Yet, since counterdiabatic terms in many-body systems are not necessarily local in space \cite{Campo12,Takahashi13,Saberi14,Sels17}, 
there are circumstances in which the STA approach is not useful for practical purposes.

Instead, a complementary bottom-up approach considers how the fidelity (termed {\it adiabatic fidelity})
between the time-evolved states and the instantaneous eigenstates 
deviates from unity under dynamical evolution.
However, for quantum many-body systems, obtaining time-evolved states and instantaneous eigenstates by solving the many-body Schr\"odinger equation 
and eigenvalue equation may be a difficult task.
It then raises the question of whether one can estimate the adiabatic fidelity without solving equations from scratch.

The approach initiated by Ref.\ \cite{Lychkovskiy17} 
and extended in Ref.\ \cite{Chen21} 
is that the adiabatic fidelity can be estimated 
by exploiting a many-body nature of the problem --- {\it generalized orthogonality catastrophe} (GOC), 
and a fundamental inequality for the evolution of unitary dynamics in Hilbert spaces --- {\it quantum speed limit} (QSL).
The GOC refers to the property wherein the overlap between instantaneous ground states and the initial ground state decays 
exponentially as both the system size and the value of the evolution parameter increase \cite{Lychkovskiy17}, 
whereas the QSL sets an intrinsic limit on how fast the time-evolved state can deviate from the initial state
\cite{Mandelstam45,Vaidman92,Pfeifer93a,Pfeifer93b,Pfeifer95,Deffner17,Gong22}.

While the estimates derived in this manner effectively capture the behavior of the actual adiabatic fidelity within the parameter space, 
an unresolved question remains, first observed numerically in Ref.\ \cite{Lychkovskiy17}:
{\it Why are the numerical values of the adiabatic fidelity and the ground state overlap nearly identical in various situations, such as when the evolution parameter is small or when the system size is large?}
The primary aim of this study is to address this question. 

Clarifying the relation between adiabatic fidelity and ground state overlap is not only of conceptual interest but also of practical value, as it enables adiabatic fidelity to be estimated from more accessible quantities without solving the full dynamics.  
Our goal is therefore to provide a mechanism-level explanation for the observed closeness and to derive quantitative bounds that render this observation useful for diagnosing adiabaticity in many-body systems.  

\subsection*{Summary of main results and contributions}

In this work, we clarify why the adiabatic fidelity and the ground state overlap often yield nearly identical values in many-body systems.  
We show that this closeness originates from the interplay of two complementary mechanisms (Sec.\ \ref{sec: a motivating limit}).  
First, when the driving parameter is small, perturbation theory shows that the time-evolved state remains closely aligned with the initial state, 
implying that adiabatic fidelity and ground state overlap coincide up to quadratic order  [Eq.\ (\ref{eq: perturbative AF and GOC})]. 
By contrast, as the evolution parameter increases, perturbative arguments lose adequacy. 
In this regime, the explanation lies in an intrinsic property of high-dimensional Hilbert spaces: vectors within the subspace complementary to the initial state tend to become nearly orthogonal to one another.  
This {\it almost-orthogonality} (Sec.\ \ref{sec: Perspective from almost orthogonality}) strongly suppresses the difference between adiabatic fidelity and ground state overlap.

To substantiate these insights, we establish reverse triangle inequalities (Sec.\ \ref{sec: Reverse triangle inequalities}) that connect adiabatic fidelity, ground state overlap, and the auxiliary overlap term [Eq.\ (\ref{eq: a more useful form for  triangle inequality})], 
thereby providing a quantitative explanation for why adiabatic fidelity and ground state overlap remain close whenever the auxiliary overlap is small---
a manifestation of the two complementary mechanisms mentioned above.
We illustrate these general findings using the driven Rice–Mele model (Sec.\ \ref{sec: Example: non-interacting Hamiltonians}), a free-fermion model amenable to partial analytical treatment; 
Analytic and numerical results confirm indistinguishability between adiabatic fidelity and ground state overlap already for system sizes of order $10^2$ [Eq.\ (\ref{eq: sDun vis final two level and k ind}) and Fig.\ \ref{Fig: F minus C}].
We further examine the robustness using the interacting Kitaev chain
 (Sec.\ \ref{sec: Example: interacting fermions}), where exact diagonalization shows that almost-orthogonality—and thus the closeness of adiabatic fidelity and ground state overlap—persists in a nonintegrable setting.

We also propose a scaling description of the auxiliary overlap in terms of the ground-state
overlap (Sec.\ \ref{sec: Improved bounds on adiabatic fidelity}) via an effective exponent $s$
[Eq.\ (\ref{eq: sD postulate form})]. This leads to a family of refined inequalities
[Eq.\ (\ref{eq: a new improved bound})] for estimating adiabatic fidelity and to a modified
adiabaticity breakdown criterion in which the critical driving rate acquires a finite multiplicative
factor [Eq.\ (\ref{eq: scaling down of Gamma general form})].

Taken together, these results demonstrate that the near-equality of adiabatic fidelity and ground state overlap is a robust and generic feature of driven many-body dynamics, 
governed by perturbative behavior at small parameters together with almost-orthogonality at large system sizes.
Beyond conceptual clarification, our results provide practical diagnostics for driven many-body dynamics.

\textbf{Relation to mathematical results on many-body adiabatic state preparation.}
Rigorous many-body adiabatic theorems provide {\it sufficient} conditions for successful adiabatic preparation, typically formulated in terms of spectral gap conditions and smooth (e.g., Gevrey-class) ramp assumptions, 
with driving rates that must decrease as system size increases (see, e.g., \cite{Jansen07,Nenciu93,Hagedorn02,Ge16,Bachmann17,Bachmann18,Bachmann19}).
Our approach is complementary: rather than offering sufficient condition guarantees under asymptotically slow ramps, we develop \emph{diagnostic}, computable criteria for adiabaticity in finite many-body systems.
We likewise conclude that the driving rate must decrease with system size to maintain adiabaticity;
in contrast, our scaling is inferred from the decay of adiabatic fidelity and ground state overlap rather than from spectral gap or Gevrey-class assumptions.
Thus, our results provide \emph{necessary} diagnostics directly applicable to concrete systems.

\textbf{Organization.}
The rest of this paper is organized as follows.  
After reviewing the basic formalism in Sec.\ \ref{sec: Preliminaries}, 
we consider a limiting case in Sec.\ \ref{sec: a motivating limit} where the essential elements for addressing the proposed question are exposed.  
We then derive a set of useful triangle-type inequalities in Sec.\ \ref{sec: Reverse triangle inequalities}.  
Our general results are illustrated using a driven Rice--Mele model in Sec.\ \ref{sec: Example: non-interacting Hamiltonians}  
and a driven interacting Kitaev chain model in Sec.\ \ref{sec: Example: interacting fermions}.  
Implications and summary are presented in Sec.\ \ref{sec: Improved bounds on adiabatic fidelity} and Sec.\ \ref{sec: Discussion and Conclusion}, respectively.  
The appendices provide additional technical details.

\section{Preliminaries}
\label{sec: Preliminaries}

\subsection{Setup}
\label{eq: setup}

We begin by defining notations and terminologies.
Consider a time-dependent Hamiltonian $H^{\,}_{\lambda}$ with $\lambda=\lambda(t)$ being an explicit function of time $t$.
Using $\lambda$ in place of $t$ as the evolution parameter,
the Schr\"odinger equation for the time-evolved state $|\Psi^{\,}_{\lambda}\rangle$ reads
\begin{align}
\mathrm{i}\Gamma\partial^{\,}_{\lambda}|\Psi^{\,}_{\lambda}\rangle
=
H^{\,}_{\lambda}|\Psi^{\,}_{\lambda}\rangle
\quad
\text{with}
\quad
|\Psi^{\,}_{0}\rangle=|\Phi^{\,}_{0}\rangle,
\label{eq: schroedinger equation}
\end{align}
where $\Gamma\:=\partial^{\,}_{t}\lambda(t)$ being the {\it driving rate}
and we assume that the initial state, $|\Psi^{\,}_{0}\rangle,$ is in the ground state of the Hamiltonian $H^{\,}_{\lambda}$ at $\lambda=0,$ $|\Phi^{\,}_{0}\rangle.$
For generic values of $\lambda,$ the instantaneous ground state of the Hamiltonian $H^{\,}_{\lambda}$ is the solution to the eigenvalue problem,
\begin{align}
H^{\,}_{\lambda}|\Phi^{\,}_{\lambda}\rangle
=
E^{\,}_{\mathrm{GS},\lambda}
|\Phi^{\,}_{\lambda}\rangle
\label{eq: instantaneous eigenvalue equation}
\end{align}
with $E^{\,}_{\mathrm{GS},\lambda}$ being the $\lambda$-dependent ground state energy.
To quantify the distance between $|\Psi^{\,}_{\lambda}\rangle$ and $|\Phi^{\,}_{\lambda}\rangle,$
one introduces the quantum fidelity $\mathcal{F}(\lambda)$ between them,
\begin{align}
\mathcal{F}(\lambda)\:=|\langle\Phi^{\,}_{\lambda}|\Psi^{\,}_{\lambda}\rangle|^2.
\label{eq: define adiabatic fidelity}
\end{align}

Let $\mathcal{C}(\lambda)$ be the overlap between the initial ground state $|\Phi^{\,}_{0}\rangle$ 
and the instantaneous ground state $|\Phi^{\,}_{\lambda}\rangle$ for an
arbitrary value of $\lambda$,
\begin{align}
\mathcal{C}(\lambda)\:=|\langle\Phi^{\,}_{\lambda}|\Phi^{\,}_{0}\rangle|^2.
\label{eq: define GOC}
\end{align}
For a large class of many-body systems, the ground state overlap $\mathcal{C}(\lambda)$ has an asymptotic form
\begin{align}
\mathcal{C}(\lambda)\sim e^{-C^{\,}_{N}\lambda^2},
\quad
C^{\,}_{N}>0,
\label{eq: scaling form of GOC}
\end{align} 
under the limit of large system size, i.e.\ $N\to\infty$ \cite{Lychkovskiy17}.  
Generalized orthogonality catastrophe, renaissance of Anderson's orthogonality catastrophe \cite{Anderson67,Gebert14},
takes place if the exponent $C^{\,}_{N}\to\infinity$ as $N\to\infinity.$
The scaling form of the exponent $C^{\,}_{N}$ depends on the type of driving, space dimensions, and whether the energy gap is present or not
\cite{Lychkovskiy17}.

At any value of $\lambda,$
we are given three vectors: $|\Phi^{\,}_{\lambda}\rangle$, $|\Psi^{\,}_{\lambda}\rangle$, and $|\Phi^{\,}_{0}\rangle.$
There are three ways to construct overlaps between any two of the three vectors.
We have already mentioned two kinds of the overlaps, namely, the adiabatic fidelity $\mathcal{F}(\lambda)$ (\ref{eq: define adiabatic fidelity}) 
and the ground state overlap $\mathcal{C}(\lambda)$ (\ref{eq: define GOC}).
The remaining overlap, $|\langle\Psi^{\,}_{\lambda}|\Phi^{\,}_{0}\rangle|^2,$
can be utilized to define the distance between the initial state $|\Phi^{\,}_{0}\rangle$ and the time-evolved state $|\Psi^{\,}_{\lambda}\rangle$
through the {\it Bures angle} $\theta(\lambda)\in[0,\pi/2]$,
\begin{align}
\theta(\lambda)\:=
\arccos|\langle\Psi^{\,}_{\lambda}|\Phi^{\,}_{0}\rangle|.
\label{eq: bures angle}
\end{align}
Given the triplet $\{\mathcal{F}(\lambda),\mathcal{C}(\lambda),\theta(\lambda)\}$, a bound on the adiabatic fidelity $\mathcal{F}(\lambda)$
around ground state overlap $\mathcal{C}(\lambda)$ was first found in Ref.\ \cite{Lychkovskiy17},
\begin{align}
\left|\mathcal{F}(\lambda) - \mathcal{C}(\lambda)\right| \leq \theta(\lambda).
\label{eq: the very old bound}
\end{align}

To facilitate practical use, we employ a Mandelstam–Tamm–type quantum speed-limit inequality
\cite{Mandelstam45,Vaidman92,Pfeifer93a,Pfeifer93b,Pfeifer95,Deffner17,Gong22},
which
sets an upper bound on the Bures angle $\theta(\lambda)$ (\ref{eq: bures angle}),
\begin{subequations}
\label{eq: bound from QSL}
\begin{align}
&\theta(\lambda)
\quad\leq\quad
\min\left(\mathcal{R}(\lambda),\frac{\pi}{2}\right)
\=:
\widetilde{\mathcal{R}}(\lambda),
\label{eq: bound from QSL a}
\\
&\text{where}
\quad
\mathcal{R}(\lambda)\:=
\int^{\lambda}_{0}
\frac{\mathrm{d}\lambda'}{|\Gamma(\lambda')|}
\sqrt{
\langle H^{2}_{\lambda'}\rangle^{\,}_{0}
-
\langle H^{\,}_{\lambda'}\rangle^{2}_{0}
},
\label{eq: bound from QSL c}
\end{align}
with $\langle\cdots\rangle^{\,}_{0}\:=\langle\Phi^{\,}_{0}|\cdots|\Phi^{\,}_{0}\rangle.$
\end{subequations}
In this work we focus on the time-dependent Hamiltonian
$H^{\,}_{\lambda}$ of the following form:
\begin{align}
H^{\,}_{\lambda}=
H^{\,}_{0}+\lambda V,
\label{eq: general form of driven Hamiltonian}
\end{align}
for which
the function $\mathcal{R}(\lambda)$ (\ref{eq: bound from QSL c}) with a positive constant driving rate $\Gamma$ reads
\begin{align}
\mathcal{R}(\lambda)=
\frac{\lambda^2}{2\Gamma}
\delta V^{\,}_{N}
\quad
\text{with}
\quad
\delta V^{\,}_{N}\:=
\sqrt{
\langle V^{2}\rangle^{\,}_{0}
-
\langle V\rangle^{2}_{0}
}.
\label{eq: bound from QSL special form}
\end{align}

\subsection{Orthogonal decomposition}
\label{sec: Orthogonal decomposition}

We now reformulate the main formalism developed in Ref.\ \cite{Chen21} using a more concise projection operator approach for later use.
Define $P=|\Phi^{\,}_{0}\rangle\langle\Phi^{\,}_{0}|$ as a projector onto the initial state
and $Q=\mathbb{I}-P$ as the complementary projector.
By definition, $P^2=P, Q^{2}=Q,$ and $PQ=QP=0.$
Consider the following orthogonal decompositions for the time-evolved state $|\Psi^{\,}_{\lambda}\rangle$ 
and the instantaneous ground state $|\Phi^{\,}_{\lambda}\rangle$,
\begin{align}
&
|\Psi^{\,}_{\lambda}\rangle=
P|\Psi^{\,}_{\lambda}\rangle
+
Q|\Psi^{\,}_{\lambda}\rangle,
\qquad
|\Phi^{\,}_{\lambda}\rangle=
P|\Phi^{\,}_{\lambda}\rangle
+
Q|\Phi^{\,}_{\lambda}\rangle.
\label{eq: expansion of states}
\end{align}
Notice that, by the construction of Eq.\ (\ref{eq: expansion of states}), the following relations hold
(here, $\||\cdot\rangle\|\:=\sqrt{\langle\cdot|\cdot\rangle}$),
\begin{subequations}
\label{eq: properties by constructions}
\begin{align}
&
\|P|\Psi^{\,}_{\lambda}\rangle\|
=
|\langle\Phi^{\,}_{0}|\Psi^{\,}_{\lambda}\rangle| 
= \cos\theta(\lambda),
\qquad
\|Q|\Psi^{\,}_{\lambda}\rangle\|
=
\sqrt{1-|\langle\Phi^{\,}_{0}|\Psi^{\,}_{\lambda}\rangle|^{2}} 
=
\sin\theta(\lambda),
\label{eq: properties by construction a}
\\
&
\|P|\Phi^{\,}_{\lambda}\rangle\|
=
|\langle\Phi^{\,}_{0}|\Phi^{\,}_{\lambda}\rangle| 
= \sqrt{C(\lambda)},
\qquad
\|Q|\Phi^{\,}_{\lambda}\rangle\|
=\sqrt{1-|\langle\Phi^{\,}_{0}|\Phi^{\,}_{\lambda}\rangle|^{2}}
=
\sqrt{1-\mathcal{C}(\lambda)},
\label{eq: properties by construction b}
\end{align}
where the Bures angle $\theta(\lambda)$ and the ground state overlap $\mathcal{C}(\lambda)$ are introduced in Eqs.\ (\ref{eq: bures angle}) 
and (\ref{eq: define GOC}), respectively.
\end{subequations}
The two vectors, $Q|\Psi^{\,}_{\lambda}\rangle$ and $Q|\Phi^{\,}_{\lambda}\rangle$, are not normalized;
we defined the corresponding normalized vectors as
\begin{align}
&|\Phi^{\perp}_{0}(\lambda)\rangle
\:=
\frac{Q|\Psi^{\,}_{\lambda}\rangle}{\|Q|\Psi^{\,}_{\lambda}\rangle\|},
\quad
|\widetilde{\Phi}^{\perp}_{0}(\lambda)\rangle
\:=
\frac{Q|\Phi^{\,}_{\lambda}\rangle}{\|Q|\Phi^{\,}_{\lambda}\rangle\|},
\label{eq: define complementary components}
\end{align}
where the superscript $\perp$ indicates that these two normalized vectors are orthogonal to the initial state $|\Phi^{\,}_{0}\rangle.$  
We introduce $\mathcal{D}(\lambda)$
to denote the overlap between the two normalized vectors, 
$|\Phi^{\perp}_{0}(\lambda)\rangle$ and $|\widetilde{\Phi}^{\perp}_{0}(\lambda)\rangle$,
\begin{align}
\mathcal{D}(\lambda)\:=&\,
|\langle \Phi^{\perp}_{0}(\lambda)|\widetilde{\Phi}^{\perp}_{0}(\lambda)\rangle|^2,
\label{eq: define overlap of orthogonal components}
\end{align}
and $\mathcal{D}^{\,}_{\mathrm{un}}(\lambda)$
to denote the overlap between the two {\it unnormalized} vectors, 
$Q|\Psi^{\,}_{\lambda}\rangle$ and $Q|\Phi^{\,}_{\lambda}\rangle$,
\begin{align}
\mathcal{D}^{\,}_{\mathrm{un}}(\lambda)
\:=
\left|
\langle\Psi^{\,}_{\lambda}|
Q
|\Phi^{\,}_{\lambda}\rangle
\right|^2.
\label{eq: define un normalized D overlap}
\end{align}
Note that the two overlaps, $\mathcal{D}(\lambda)$ 
and $\mathcal{D}^{\,}_{\mathrm{un}}(\lambda)$,
are not independent.
They are related through 
\begin{align}
\sqrt{\mathcal{D}^{\,}_{\mathrm{un}}(\lambda)}
&=
\sin\theta(\lambda)
\sqrt{
1-\mathcal{C}(\lambda)
}
\sqrt{\mathcal{D}(\lambda)}.
\label{eq: relation between two D overlap}
\end{align}
Both the normalized overlap $\mathcal{D}(\lambda)$ and the unnormalized overlap $\mathcal{D}^{\,}_{\mathrm{un}}(\lambda)$
play important roles in the following discussion.

It was found in 
Ref.\ \cite{Chen21} 
(for an alternative derivation using the projection–operator formalism, see Appendix~\ref{App0: Perturbative derivation of inequality})
that the difference between the adiabatic fidelity $\mathcal{F}(\lambda)$ (\ref{eq: define adiabatic fidelity}) 
and the ground state overlap
$\mathcal{C}(\lambda)$ (\ref{eq: define GOC})
obeys the following inequality
\begin{align}
\left|
\mathcal{F}(\lambda)
-
\mathcal{C}(\lambda)
\right|
\quad\leq\quad
\left|
-
\sin^2\theta(\lambda)
\mathcal{C}(\lambda)
+
\mathcal{D}^{\,}_{\mathrm{un}}(\lambda)
\right|
+
2
\cos\theta(\lambda)
\sqrt{\mathcal{C}(\lambda)}
\sqrt{\mathcal{D}^{\,}_{\mathrm{un}}(\lambda)},
\label{eq: inequality perp basis}
\end{align}
where $\mathcal{D}^{\,}_{\mathrm{un}}(\lambda)$ is defined in 
Eq.\ (\ref{eq: define un normalized D overlap}).
To make further progress, the strategy made in Ref.\ \cite{Chen21} was to replace the normalized overlap 
$\mathcal{D}(\lambda)$ of Eq.\
(\ref{eq: relation between two D overlap})
by its trivial upper bound 1, 
i.e., $\mathcal{D}(\lambda)
\leq1,$
which renders the unnormalized overlap 
$\mathcal{D}^{\,}_{\mathrm{un}}(\lambda)$ (\ref{eq: relation between two D overlap})
bounded from above as follows
\begin{align}
\sqrt{\mathcal{D}^{\,}_{\mathrm{un}}(\lambda)}
\quad\leq\quad
\sin^{\,}\theta(\lambda)
\sqrt{1-\mathcal{C}(\lambda)}.
\label{eq: trivial upper bound for fidelity}
\end{align}
The rationale for adopting the trivial upper bound, $\mathcal{D}(\lambda)\leq1,$
is that, since presumably we have no knowledge about the overlap between 
the two normalized vectors, $|\Phi^{\perp}_{0}(\lambda)\rangle$ and $|\widetilde{\Phi}^{\perp}_{0}(\lambda)\rangle$
(\ref{eq: define complementary components}),
we may simply replace their overlap with the trivial upper bound of their overlap.
Applying the upper bound (\ref{eq: trivial upper bound for fidelity}) to
the inequality (\ref{eq: inequality perp basis}) yields
\begin{align}
\left|\mathcal{F}(\lambda)-\mathcal{C}(\lambda)\right|
\quad\leq\quad
\sin^2\theta
\left|
1-
2\mathcal{C}
\right|
+
\sin(2\theta)
\sqrt{\mathcal{C}}\sqrt{1-\mathcal{C}}.
\label{eq: define an auxiliary function}
\end{align}

Although the inequality (\ref{eq: define an auxiliary function}) offers an improvement over the inequality (\ref{eq: the very old bound})
found in Ref.\ \cite{Lychkovskiy17}, 
it remains unclear why the values of the adiabatic fidelity $\mathcal{F}(\lambda)$ and the ground state overlap $\mathcal{C}(\lambda)$ are nearly identical when (i) the system size $N$ is sufficiently large (e.g., $N \geq 100),$ or (ii) the evolution parameter $\lambda$ is small for any system size.
In the present work, we address this question using the orthogonal decomposition formalism detailed in this section.

\section{A motivating limit and interpretations}
\label{sec: a motivating limit}

By inspecting Eq.\ 
(\ref{eq: inequality perp basis}),
one observes that the least controlled piece in the inequality is the unnormalized overlap $\mathcal{D}^{\,}_{\mathrm{un}}(\lambda)$ (\ref{eq: define un normalized D overlap}),
which contains two factors [see Eq.\ (\ref{eq: relation between two D overlap})], namely, 
$\sqrt{D(\lambda)}$ and $\sin\theta(\lambda)\sqrt{1-\mathcal{C}(\lambda)}$.
Among them, 
the trivial upper bound of $\mathcal{D}(\lambda)$ is employed in Ref.\ \cite{Chen21} to 
obtain universal upper bounds on $|\mathcal{F}(\lambda)-\mathcal{C}(\lambda)|$ [see Eq.\ (\ref{eq: define an auxiliary function})].
Therefore, the reason why the inequality (\ref{eq: define an auxiliary function}) is insufficient to explain the smallness of
$|\mathcal{F}(\lambda)-\mathcal{C}(\lambda)|$ 
may stem from the use of the trivial upper bound, $\mathcal{D}(\lambda)\leq1.$ 
To justify this claim, we simply look at the 
extreme limit:
\begin{align}
\mathcal{D}^{\,}_{\mathrm{un}}(\lambda)
\to
0.
\label{eq: orhogonality of orthogonal complements}
\end{align}
We will further elaborate on the orthogonality limit (\ref{eq: orhogonality of orthogonal complements}) later. 
For now, let us examine its consequences.
Imposing the orthogonality limit (\ref{eq: orhogonality of orthogonal complements}) to the defining equations (\ref{eq: expansion of states}),
the calculation of $|\mathcal{F}(\lambda)-\mathcal{C}(\lambda)|$ is fairly simple.
First, we find from Eqs.\ (\ref{eq: expansion of states}) and (\ref{eq: properties by constructions}) that 
\begin{align}
\mathcal{F}(\lambda)
\to
\cos^2\theta(\lambda)
\mathcal{C}(\lambda).
\label{eq: inequality when exact orthogonality holds final result F form}
\end{align}
It then follows that
\begin{align}
|\mathcal{F}(\lambda)-\mathcal{C}(\lambda)|
\to
\sin^2\theta(\lambda)\,\mathcal{C}(\lambda).
\label{eq: inequality when exact orthogonality holds final result}
\end{align}
Comparing the right side of Eq.\ (\ref{eq: inequality when exact orthogonality holds final result}) 
with that of (\ref{eq: define an auxiliary function}) 
indicates that only a portion of Eq.\ (\ref{eq: define an auxiliary function}) is retained on the right side of 
Eq.\ (\ref{eq: inequality when exact orthogonality holds final result}), resulting in a stronger upper bound on $|\mathcal{F}(\lambda)-\mathcal{C}(\lambda)|.$

The orthogonality limit (\ref{eq: orhogonality of orthogonal complements}), in view of Eq.\ (\ref{eq: relation between two D overlap}), 
can be achieved by either
\begin{align}
\text{(i)}\;
\sin\theta(\lambda)\sqrt{1-\mathcal{C}(\lambda)}
\to
0 
\quad 
\text{or}
\quad
\text{(ii)}\;
\sqrt{D(\lambda)}
\to
0.
\label{eq: two cases}
\end{align}
Case (i) is satisfied if $\lambda$ is small.
This is anticipated since if $\lambda$ is small, one expects that the Bures angle $\theta(\lambda)$ 
is still small as well, while both $\mathcal{C}(\lambda)$ and $\mathcal{D}(\lambda)$ remain close to one. 
To support this argument, we will present 
in Sec.\ \ref{sec: Perspective from perturbation theory} 
an explicit calculation based on the perturbative expansion in $\lambda$.
As for case (ii), we shall see  
in Sec.\ \ref{sec: Perspective from almost orthogonality} 
that it can be understood as a manifestation of almost-orthogonality occurring in the complementary subspace of the initial state 
$|\Phi^{\,}_{0}\rangle.$
We now elaborate on the two cases of Eq.\ (\ref{eq: two cases}) in turn.

\subsection{Insights from perturbative expansion in $\lambda$}
\label{sec: Perspective from perturbation theory}

Here, we provide a further explanation for case (i) of Eq.\ (\ref{eq: two cases}).
Our objective is to solve the instantaneous eigenvalue equation (\ref{eq: instantaneous eigenvalue equation}) 
and the time-dependent Schr\"odinger equation (\ref{eq: schroedinger equation}) perturbatively in $\lambda$
for the Hamiltonian $H^{\,}_{\lambda}$ presented in Eq.\ (\ref{eq: general form of driven Hamiltonian}),
given the eigenvalue equation of $H^{\,}_{0},$
$
H^{\,}_{0}
|\chi^{\,}_{n}\rangle
=
\varepsilon^{\,}_{n}|\chi^{\,}_{n}\rangle,
$
where $\{|\chi^{\,}_{n}\rangle\}$ is a complete set of orthonormal eigenstates of $H^{\,}_{0}$ with $|\chi^{\,}_{0}\rangle\equiv|\Psi^{\,}_{0}\rangle$ being its ground state
and $n\in\{0,1,\cdots\}$ labels different eigenstates.
One finds (refer to Appendix \ref{App1: Perturbative  expansion} for details) that, up to order $\lambda^2,$ 
the adiabatic fidelity $\mathcal{F}(\lambda)$ (\ref{eq: define adiabatic fidelity}) and the ground state overlap $\mathcal{C}(\lambda)$ 
(\ref{eq: define GOC})
are identical and are independent of the driving rate $\Gamma$,
\begin{align}
\mathcal{F}(\lambda)
\simeq
\mathcal{C}(\lambda)
=
1
-
\lambda^{2}
\sum^{\,}_{n\neq 0}
\frac{|V^{\,}_{n0}|^2}{(\varepsilon^{\,}_{0}-\varepsilon^{\,}_{n})^2}
+
\mathcal{O}(\lambda^3),
\label{eq: perturbative AF and GOC}
\end{align}
where the matrix element $V^{\,}_{nm}\:=\langle\chi^{\,}_{n}|V|\chi^{\,}_{m}\rangle.$
The difference between 
$\mathcal{F}(\lambda)$
and 
$\mathcal{C}(\lambda)$
appears at order $\lambda^3,$
$
\mathcal{F}(\lambda)-\mathcal{C}(\lambda)=
-
\lambda^{3}
V^{\,}_{00}\varepsilon^{\,}_{0}/\Gamma^2
+\cdots.
$
Leading order contributions for various quantities
can also be obtained,
\begin{subequations}
\label{eq: perturbative other quantities}
\begin{align}
&
\sin\theta(\lambda)
=
\frac{\lambda^2}{2\Gamma}
\Big(
\sum^{\,}_{n\neq0}
|V^{\,}_{n0}|^2
\Big)^{1/2}
+\mathcal{O}(\lambda^3),
\label{eq: perturbative other quantities a}
\\
&
\sqrt{\mathcal{D}^{\,}_{\mathrm{un}}(\lambda)}
=
\frac{\lambda^3}{2\Gamma}
\sum^{\,}_{n\neq0}
\frac{|V^{\,}_{n0}|^2}{\varepsilon^{\,}_{n}-\varepsilon^{\,}_{0}}
+
\mathcal{O}(\lambda^4),
\label{eq: perturbative other quantities b}
\\
&
\sin\theta(\lambda)\sqrt{1-\mathcal{C}(\lambda)}
=
\mathcal{O}(\lambda^3),
\label{eq: perturbative other quantities c}
\\
&
\sqrt{\mathcal{D}(\lambda)}
=
\mathcal{O}(1).
\label{eq: perturbative other quantities d}
\end{align}
\end{subequations}
We see that, for small $\lambda$, $\sqrt{\mathcal{D}^{\,}_{\mathrm{un}}(\lambda)}$ (\ref{eq: relation between two D overlap}) 
is of order $\lambda^{3},$
which is attributed to the same order of small $\sin\theta(\lambda)\sqrt{1-\mathcal{C}(\lambda)}$ 
(\ref{eq: perturbative other quantities c}) since $\sqrt{\mathcal{D}(\lambda)}$ (\ref{eq: perturbative other quantities d}) is of order one.
However, as $\lambda$ continues to increase, the result from perturbation theory 
is insufficient to explain the smallness of 
$\sqrt{\mathcal{D}^{\,}_{\mathrm{un}}(\lambda)}.$
Instead, when $\lambda$ is not small, the almost-orthogonality exhibited in the normalized overlap $\sqrt{\mathcal{D}(\lambda)}$,
as shown in case (ii) of Eq.\ (\ref{eq: two cases}),
should be taken into account.

\subsection{Insights from almost-orthogonality in the complementary subspace under large system size}
\label{sec: Perspective from almost orthogonality}

Case (ii) of Eq.\ (\ref{eq: two cases})
may be understood as follows.
Let 
$
\left\{|\Phi^{\,}_{0}\rangle, |u^{\,}_{1}\rangle,|u^{\,}_{2}\rangle,\cdots,|u^{\,}_{\mathfrak{n}-1}\rangle\right\}
$
be a complete set of $\lambda$-independent orthonormal basis in an $\mathfrak{n}$-dimensional Hilbert space $\mathscr{H}^{\,}_{\mathfrak{n}}$.
Since both the time-evolved state $|\Psi^{\,}_{\lambda}\rangle$ and the instantaneous ground state $|\Phi^{\,}_{\lambda}\rangle$ 
are vectors in the full Hilbert space $\mathscr{H}^{\,}_{\mathfrak{n}}$,
it follows from the orthogonal decomposition (\ref{eq: expansion of states}) that 
the two normalized vectors, 
$|\Phi^{\perp}_{0}(\lambda)\rangle$ 
and 
$|\widetilde{\Phi}^{\perp}_{0}(\lambda)\rangle$ (\ref{eq: define complementary components}),
are vectors lying in the subspace $\mathscr{H}^{\perp}_{\mathfrak{n}-1}$, 
where the codimension-1 Hilbert space $\mathscr{H}^{\perp}_{\mathfrak{n}-1}$
is spanned by 
$
\{|u^{\,}_{1}\rangle,|u^{\,}_{2}\rangle,\cdots,|u^{\,}_{\mathfrak{n}-1}\rangle\}.
$
When $\mathfrak{n}$ is large, 
the two normalized vectors, $|\Phi^{\perp}_{0}(\lambda)\rangle$ and $|\widetilde{\Phi}^{\perp}_{0}(\lambda)\rangle$, 
may be thought of as two independent {\it random vectors} in the Hilbert space $\mathscr{H}^{\perp}_{\mathfrak{n}-1}$
even though the vector $|\Phi^{\perp}_{0}(\lambda)\rangle$ undergoes dynamical evolution  
while the other vector
$|\widetilde{\Phi}^{\perp}_{0}(\lambda)\rangle$ experiences adiabatic transformation.
As a result, one would expect their overlap, $\mathcal{D}(\lambda)$ (\ref{eq: define overlap of orthogonal components}), 
to decay sufficiently fast with increasing $\mathfrak{n}$.
Following literature in mathematics \cite{Vershynin18}, we refer to this kind of orthogonal property as {\it almost-orthogonality}. 

\begin{figure}[t]
\begin{center}
\includegraphics[width=0.2\textwidth]{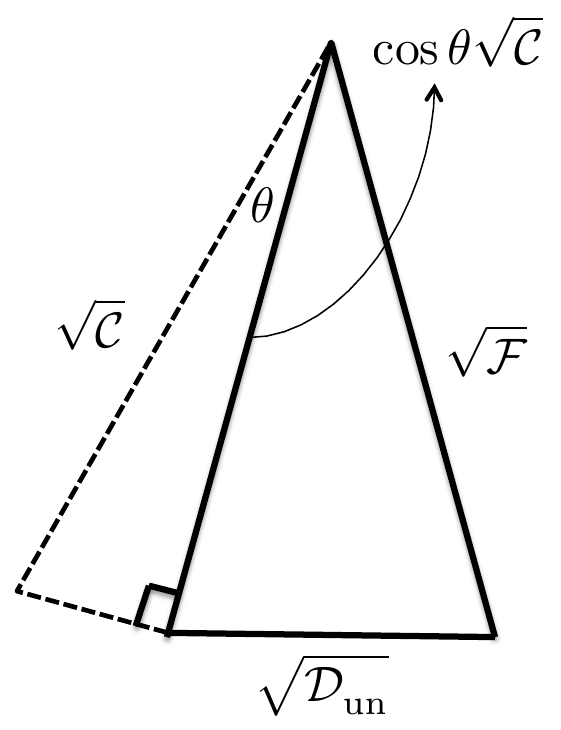}
\caption{
Depict of the triangle relationship (\ref{eq: reverse triangle inequalities}) between the three real-valued quantities:
$\sqrt{\mathcal{F}(\lambda)}$ (\ref{eq: define adiabatic fidelity}),
$\cos\theta(\lambda)\,\sqrt{\mathcal{C}(\lambda)}$ (\ref{eq: define GOC}), 
and $\sqrt{\mathcal{D}^{\,}_{\mathrm{un}}(\lambda)}$ (\ref{eq: define un normalized D overlap}),
where $\theta(\lambda)$ is the Bures angle (\ref{eq: bures angle}).
\label{Fig: triangle relation}
         }
\end{center}
\end{figure}

\section{Reverse triangle inequalities}
\label{sec: Reverse triangle inequalities}

The discussion presented in the above section [see Eq.\ (\ref{eq: inequality when exact orthogonality holds final result})]
indicates that $\mathcal{F}(\lambda)$ is identical to $\cos^{2}\theta(\lambda)\mathcal{C}(\lambda)$
under the exact orthogonality limit (\ref{eq: orhogonality of orthogonal complements}).
This observation motivates us to pursue bounds on the difference between them, 
namely, $|\sqrt{\mathcal{F}(\lambda)}-\cos^{\,}\theta(\lambda)\sqrt{\mathcal{C}(\lambda)}|$.
A useful tool for the present work is the following lemma.
\begin{lemmanon}
The three real-valued quantities, 
$\sqrt{\mathcal{F}(\lambda)}$ (\ref{eq: define adiabatic fidelity}),
$\cos\theta(\lambda)\,\sqrt{\mathcal{C}(\lambda)}$ (\ref{eq: define GOC}), 
and $\sqrt{\mathcal{D}^{\,}_{\mathrm{un}}(\lambda)}$ (\ref{eq: define un normalized D overlap}),
obey a set of (reverse) triangle inequalities,
\begin{subequations}
\label{eq: reverse triangle inequalities} 
\begin{align}
|\sqrt{\mathcal{F}(\lambda)}-\cos^{\,}\theta(\lambda)\sqrt{\mathcal{C}(\lambda)}|
&\quad\leq\quad
\sqrt{\mathcal{D}^{\,}_{\mathrm{un}}(\lambda)}.
\label{eq: reverse triangle inequalities 1}  
\\
\left|
\sqrt{\mathcal{D}^{\,}_{\mathrm{un}}(\lambda)}
-
\cos\theta(\lambda)\,\sqrt{\mathcal{C}(\lambda)}
\right|
&\quad\leq\quad
\sqrt{\mathcal{F}(\lambda)},
\label{eq: reverse triangle inequalities 2} 
\\
\left|
\sqrt{\mathcal{D}^{\,}_{\mathrm{un}}(\lambda)}
-
\sqrt{\mathcal{F}(\lambda)}
\right|
&\quad\leq\quad
\cos\theta(\lambda)\,\sqrt{\mathcal{C}(\lambda)}.
\label{eq: reverse triangle inequalities 3} 
\end{align}
\end{subequations}
\end{lemmanon}
Thus, 
$\sqrt{\mathcal{F}(\lambda)}$,
$\cos\theta(\lambda)\,\sqrt{\mathcal{C}(\lambda)}$,
and $\sqrt{\mathcal{D}^{\,}_{\mathrm{un}}(\lambda)}$, 
form a triangle on a plane for all values of parameters. See Fig.\ \ref{Fig: triangle relation} for an illustration.
\begin{proof}
First, we begin by considering the right side of Eq.\ (\ref{eq: reverse triangle inequalities 1}) with the help of Eq.\ (\ref{eq: define un normalized D overlap}),
\begin{align}
\sqrt{\mathcal{D}^{\,}_{\mathrm{un}}(\lambda)}
&=
\Big|
\langle\Psi^{\,}_{\lambda}|\Phi^{\,}_{\lambda}\rangle
-
\langle\Psi^{\,}_{\lambda}|\Phi^{\,}_{0}\rangle\langle\Phi^{\,}_{0}|\Phi^{\,}_{\lambda}\rangle
\Big|
\nonumber\\
&\geq
\Big|
|\langle\Psi^{\,}_{\lambda}|\Phi^{\,}_{\lambda}\rangle|
-
|\langle\Psi^{\,}_{\lambda}|\Phi^{\,}_{0}\rangle||\langle\Phi^{\,}_{0}|\Phi^{\,}_{\lambda}\rangle|
\Big|
=
\Big|
\sqrt{\mathcal{F}(\lambda)}
-
\cos\theta(\lambda)\,
\sqrt{\mathcal{C}(\lambda)}
\Big|,
\label{eq: reverse triangle inequalities 1 pf} 
\end{align}
where we have used the reverse triangle inequality $|z-w|\geq\left||z|-|w|\right|$ for $z,w\in\mathbb{C}.$
The inequality (\ref{eq: reverse triangle inequalities 1}) is thus established.
Note that the right side of Eq.\ (\ref{eq: reverse triangle inequalities 1 pf}) 
may be interpreted as a lower bound on $\sqrt{\mathcal{D}^{\,}_{\mathrm{un}}(\lambda)}.$

Second, we observe that there is an upper bound on $\sqrt{\mathcal{D}^{\,}_{\mathrm{un}}(\lambda)},$
\begin{align}
\sqrt{\mathcal{D}^{\,}_{\mathrm{un}}(\lambda)}
&=
\Big|
\langle\Psi^{\,}_{\lambda}|\Phi^{\,}_{\lambda}\rangle
-
\langle\Psi^{\,}_{\lambda}|\Phi^{\,}_{0}\rangle\langle\Phi^{\,}_{0}|\Phi^{\,}_{\lambda}\rangle
\Big|
\nonumber\\
&\leq
|\langle\Psi^{\,}_{\lambda}|\Phi^{\,}_{\lambda}\rangle|
+
|\langle\Psi^{\,}_{\lambda}|\Phi^{\,}_{0}\rangle||\langle\Phi^{\,}_{0}|\Phi^{\,}_{\lambda}\rangle|
=
\sqrt{\mathcal{F}(\lambda)}
+
\cos\theta(\lambda)\,
\sqrt{\mathcal{C}(\lambda)},
\label{eq: upper bound on Dun}
\end{align}
which is a consequence of the triangle inequality, $|z+w|\leq|z|+|w|$ for $z,w\in\mathbb{C}.$

Finally, 
combing the upper bound (\ref{eq: upper bound on Dun}) with the lower bound (\ref{eq: reverse triangle inequalities 1 pf}) yields two two-sided bounds
on $\sqrt{\mathcal{D}^{\,}_{\mathrm{un}}(\lambda)}$ (neglecting $\lambda$ to simplify notation),
\begin{subequations}
\begin{align}
-\sqrt{\mathcal{F}}+\cos\theta\,\sqrt{\mathcal{C}}
&\quad\leq\quad
\sqrt{\mathcal{D}^{\,}_{\mathrm{un}}}
\quad\leq\quad
\sqrt{\mathcal{F}}+\cos\theta\,\sqrt{\mathcal{C}},
\\
\sqrt{\mathcal{F}}-\cos\theta\,\sqrt{\mathcal{C}}
&\quad\leq\quad
\sqrt{\mathcal{D}^{\,}_{\mathrm{un}}}
\quad\leq\quad
\sqrt{\mathcal{F}}+\cos\theta\,\sqrt{\mathcal{C}}.
\end{align}
\end{subequations}
This completes the proof of Eqs.\ (\ref{eq: reverse triangle inequalities 2}) and (\ref{eq: reverse triangle inequalities 3}).
\end{proof}

The first triangle inequality (\ref{eq: reverse triangle inequalities 1})
provides a quantitative way to understand the closeness between $\mathcal{F}(\lambda)$ and $\mathcal{C}(\lambda)$ since
\begin{align}
\sqrt{\mathcal{F}}
-
\sqrt{\mathcal{C}}
\quad\leq\quad
\sqrt{\mathcal{F}}
-
\cos\theta(\lambda)\sqrt{\mathcal{C}}
\quad\leq\quad
\sqrt{\mathcal{D}^{\,}_{\mathrm{un}}}.
\label{eq: a more useful form for  triangle inequality}
\end{align}
Given that the unnormalized overlap $\sqrt{\mathcal{D}^{\,}_{\mathrm{un}}(\lambda)}$ acts as an upper bound,
a small value for it,
which can be achieved by the two cases of Eq.\ (\ref{eq: two cases}),
suggests that the numerical difference between the adiabatic fidelity $\mathcal{F}(\lambda)$ and the ground state overlap $\mathcal{C}(\lambda)$ 
must be even smaller.

\section{Illustrative example I: non-interacting Hamiltonians}
\label{sec: Example: non-interacting Hamiltonians}

To illustrate our general analytical findings from Secs.\ \ref{sec: a motivating limit} and \ref{sec: Reverse triangle inequalities}, 
the remaining task is to explicitly express $\sqrt{\mathcal{D}^{\,}_{\mathrm{un}}(\lambda)}$, as defined in Eq.\ (\ref{eq: define un normalized D overlap}), 
in terms of the Bures angle $\theta(\lambda)$ 
and the ground state overlap $\mathcal{C}(\lambda)$ for specific models.
This can be done analytically for non-interacting Hamiltonians for which one obtains [see Appendix \ref{App2: Non-interacting Hamiltonians}],
\begin{align}
&
\sqrt{\mathcal{D}^{\,}_{\mathrm{un}}(\lambda)}
\simeq
\cos\theta(\lambda)
\sqrt{\mathcal{C}(\lambda)}
\left|
\sum^{\,}_{k}
A^{\,}_{k}
\right|,
\qquad
\text{where}
\quad
A^{\,}_{k}\:=
\frac{
\langle\psi^{\,}_{\lambda}(k)|(\mathbb{I}^{\,}_{k}-p^{\,}_{k})|\phi^{\,}_{\lambda}(k)\rangle
}{\langle\psi^{\,}_{\lambda}(k)|p^{\,}_{k}|\phi^{\,}_{\lambda}(k)\rangle}.
\label{eq: general form of Dun}
\end{align}
Here, $|\psi^{\,}_{\lambda}(k)\rangle,$ $|\phi^{\,}_{\lambda}(k)\rangle$, and $p^{\,}_{k}$
are the single-body counterparts of
$|\Psi^{\,}_{\lambda}\rangle,$ $|\Phi^{\,}_{\lambda}\rangle$, and $P$ introduced in Sec.\ \ref{sec: Preliminaries}, respectively.

\begin{figure}[t]
\begin{center}
\includegraphics[width=0.26\textwidth]{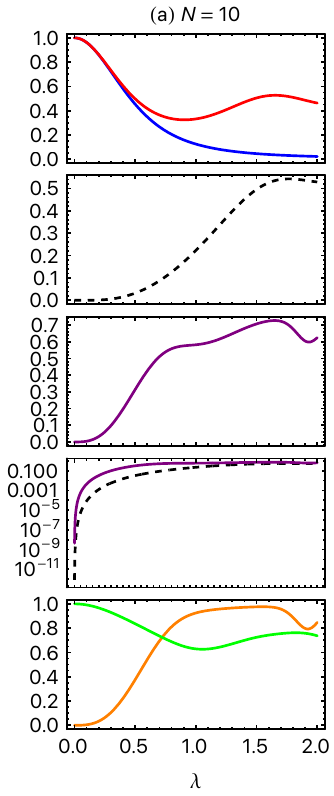}
\includegraphics[width=0.26\textwidth]{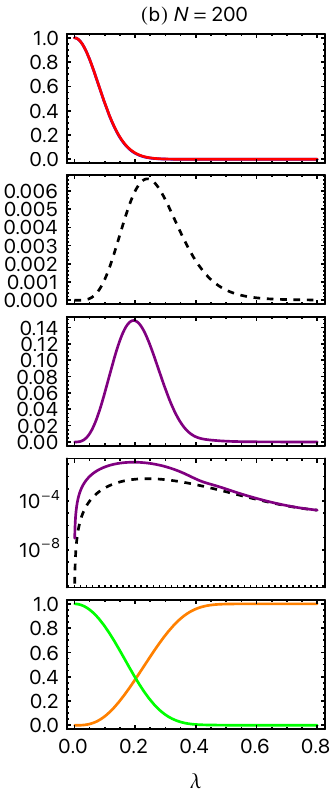}
\includegraphics[width=0.455\textwidth]{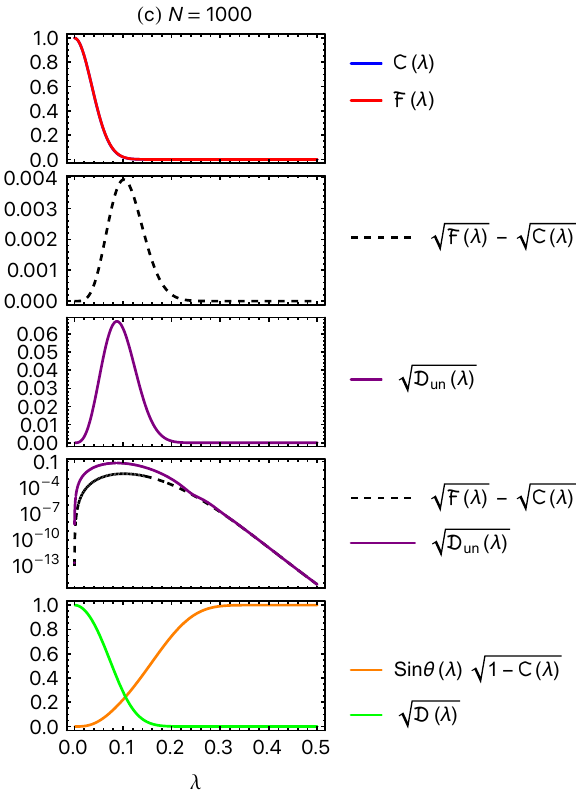}
\caption{
(Color online)
Various quantities are calculated numerically for the driven Rice-Mele model (\ref{eq: driven RM model})  
with the value of parameters shown in Eq.\ (\ref{eq: numerical parameter})
for system size $N=10,200$ and $1000.$
The fourth row overlays the curves from the second and third rows, with the vertical axis on a logarithmic scale.
Further explanation is provided in the main text.
\label{Fig: F minus C}
         }
\end{center}
\end{figure}

For concreteness, let us
consider a time-dependent Rice-Mele model describing a system of fermions 
on a half-filled one-dimensional bipartite lattice with the Hamiltonian
\cite{Rice82,Nakajima16,Lychkovskiy17}
\begin{align}
H^{\,}_{\mathrm{RM}}\:=&
\sum^{N}_{j=1}
\left[
-(J+U)a^{\dag}_{j}b^{\,}_{j}
-(J-U)a^{\dag}_{j}b^{\,}_{j+1}
+
\mathrm{h.c.}
\right]
+
\sum^{N}_{j=1}
\mu(\lambda)
\left(
a^{\dag}_{j}
a^{\,}_{j}
-
b^{\dag}_{j}
b^{\,}_{j}
\right),
\label{eq: driven RM model}
\end{align}
where $N$, the number of lattice sites, is assumed to be even.
Here, $a^{\,}_{j}$ and $b^{\,}_{j}$ are the fermionic annihilation operators on the $a$ and $b$ sublattices, respectively.
For this model with $\mu(\lambda)=\lambda$, the ground state overlap $\mathcal{C}(\lambda)$ (\ref{eq: define GOC})
and the function $\mathcal{R}(\lambda)$ (\ref{eq: bound from QSL special form}) 
take the form shown in 
Eqs.\ (\ref{eq: scaling form of GOC}) and (\ref{eq: bound from QSL special form})
with $C^{\,}_{N}=(16JU)^{-1}N$ and $\delta V^{\,}_{N}=\sqrt{N}.$

We shall specialize to the case where $J=U=$ constant
for which the summation in Eq.\ (\ref{eq: general form of Dun}) 
can be evaluated in closed form
[see Appendix \ref{App2: Non-interacting Hamiltonians}],
\begin{subequations}
\label{eq: sDun vis final two level and k ind}
\begin{align}
\sqrt{\mathcal{D}^{\,}_{\mathrm{un}}(\lambda)}
\simeq&\,
\sqrt{\mathcal{C}(\lambda)}
\cos\theta(\lambda)
\sin\theta(\lambda)
\alpha(\lambda),
\label{eq: sDun vis final two level and k ind a}
\\
\sqrt{\mathcal{D}(\lambda)}
\simeq&\,
\sqrt{\mathcal{C}(\lambda)}
\cos\theta(\lambda)
\alpha(\lambda)
/
\sqrt{
1-\mathcal{C}(\lambda)
},
\label{eq: D overlap analytical form}
\end{align}
where we have introduced an auxiliary function $\alpha(\lambda)$ for later convenience
\begin{align}
\alpha(\lambda)\:=&\,
\sqrt{N}\sqrt{1-\mathcal{C}(\lambda)^{1/N}}.
\label{eq: sDun vis final two level and k ind b}
\end{align}
\end{subequations}
We shall examine whether the explicit form of $\sqrt{\mathcal{D}^{\,}_{\mathrm{un}}(\lambda)}$ 
(\ref{eq: sDun vis final two level and k ind a}) 
has the desired characteristics under the limit of small $\lambda$ or large $N$
as claimed previously in Sec.\ \ref{sec: a motivating limit}.
First,  
it is readily checked that, for small $\lambda$,
the leading order contribution to $\sqrt{\mathcal{D}^{\,}_{\mathrm{un}}(\lambda)}$ (\ref{eq: sDun vis final two level and k ind a})
and $\sqrt{\mathcal{D}(\lambda)}$ (\ref{eq: D overlap analytical form})
are $\mathcal{O}(\lambda^3)$ and $\mathcal{O}(1)$, respectively,
which is consistent with the results obtained from general perturbation theory as presented in Eq.\ (\ref{eq: perturbative other quantities}).
Second, note that the auxiliary function $\alpha(\lambda)$ (\ref{eq: sDun vis final two level and k ind b}) scales at most with a rate of $\sqrt{N}$:
\begin{align}
\alpha(\lambda)\leq \sqrt{cN\lambda^{2}}=\sqrt{-\ln\mathcal{C}(\lambda)},
\end{align}
as a result of the inequality $1-e^{-x}\leq x$ for all $x\in\mathbb{R}.$
Therefore, we deduce from Eq.\ (\ref{eq: sDun vis final two level and k ind}) that 
$
\sqrt{\mathcal{D}^{\,}_{\mathrm{un}}(\lambda)}
\sim
\sqrt{\mathcal{D}(\lambda)}
\sim
\sqrt{\mathcal{C}(\lambda)}\sqrt{N}\to0
$
as $N\to\infty,$
which is in agreement with the case (ii) of Eq.\ (\ref{eq: two cases}).

To numerically demonstrate our findings, we choose the following value of parameters
\begin{align}
(J,U,\Gamma)=(0.4,0.4,0.7)
\label{eq: numerical parameter}
\end{align} 
in the Hamiltonian $H^{\,}_{\mathrm{RM}}$ (\ref{eq: driven RM model})
as a representative example.
In Fig.\ \ref{Fig: F minus C},
we plot various quantities for the driven Rice-Mele model (\ref{eq: driven RM model})  
with system size $N=10, 200$, and $1000.$
In the first row, the adiabatic fidelity $\mathcal{F}(\lambda)$ and the ground state overlap $\mathcal{C}(\lambda)$ are indistinguishable for $N=200$ and $N=1000$.
The second row shows that, 
for both $N=200$ and $N=1000$, the difference between $\sqrt{\mathcal{F}(\lambda)}$ and $\sqrt{\mathcal{C}(\lambda)}$ raises as $\lambda$ increases 
and then diminishes as $\lambda$ further increases. 
This bell-shaped  curve of $\sqrt{\mathcal{F}(\lambda)}-\sqrt{\mathcal{C}(\lambda)}$ is in phase with the curve of 
the unnormalized overlap $\sqrt{\mathcal{D^{\,}_{\mathrm{un}}}(\lambda)}$ 
[third row].
This is consistent with Eq.~(\ref{eq: a more useful form for  triangle inequality}),
which states that
\(\sqrt{\mathcal{F}(\lambda)}-\sqrt{\mathcal{C}(\lambda)}
\le \sqrt{\mathcal{D}_{\mathrm{un}}(\lambda)}\) (see also the fourth row).
In the fourth row, we overlay
\(\sqrt{\mathcal{F}(\lambda)}-\sqrt{\mathcal{C}(\lambda)}\) (dashed) and
\(\sqrt{\mathcal{D}_{\mathrm{un}}(\lambda)}\) (purple) in a single panel with
the vertical axis on a logarithmic scale, making their separation visible.
Since $\sqrt{\mathcal{D^{\,}_{\mathrm{un}}}(\lambda)}$ can be factorized into two pieces, c.f.\ Eq.\ (\ref{eq: relation between two D overlap}),
the smallness of the monotonically {\it increasing} part of $\sqrt{\mathcal{F}(\lambda)}-\sqrt{\mathcal{C}(\lambda)}$ 
is attributed to the smallness of $\sin\theta(\lambda)\sqrt{1-\mathcal{C}(\lambda)}$ [fifth row].
Likewise, the monotonically {\it decreasing} part of $\sqrt{\mathcal{F}(\lambda)}-\sqrt{\mathcal{C}(\lambda)}$ is particularly small 
due to the almost-orthogonality occurring in the complementary space when $N$ is large, which is manifested by a small 
$\sqrt{\mathcal{D}(\lambda)}$ [fifth row].
By contrast, 
for $N=10$ [see Fig.\ \ref{Fig: F minus C}(a)], $\sqrt{\mathcal{F}(\lambda)}-\sqrt{\mathcal{C}(\lambda)}$ [second row of panel (a)] 
is monotonically increasing in most of the values of $\lambda$
and is small only in the region of small $\lambda$ (say, $\lambda\leq0.2$).
Again, this smallness of $\sqrt{\mathcal{F}(\lambda)}-\sqrt{\mathcal{C}(\lambda)}$ is related to the smallness of $\sin\theta(\lambda)\sqrt{1-\mathcal{C}(\lambda)}$ 
[fifth row of panel (a)].
When $\lambda$ further increases, however, the difference between $\sqrt{\mathcal{F}(\lambda)}$ and $\sqrt{\mathcal{C}(\lambda)}$ is notable
since the normalized overlap $\sqrt{\mathcal{D}(\lambda)}$ [fifth row of panel (a)] does not exhibit almost-orthogonality for $N=10$.

\begin{figure}[t]
\begin{center}
\includegraphics[width=0.7\textwidth]{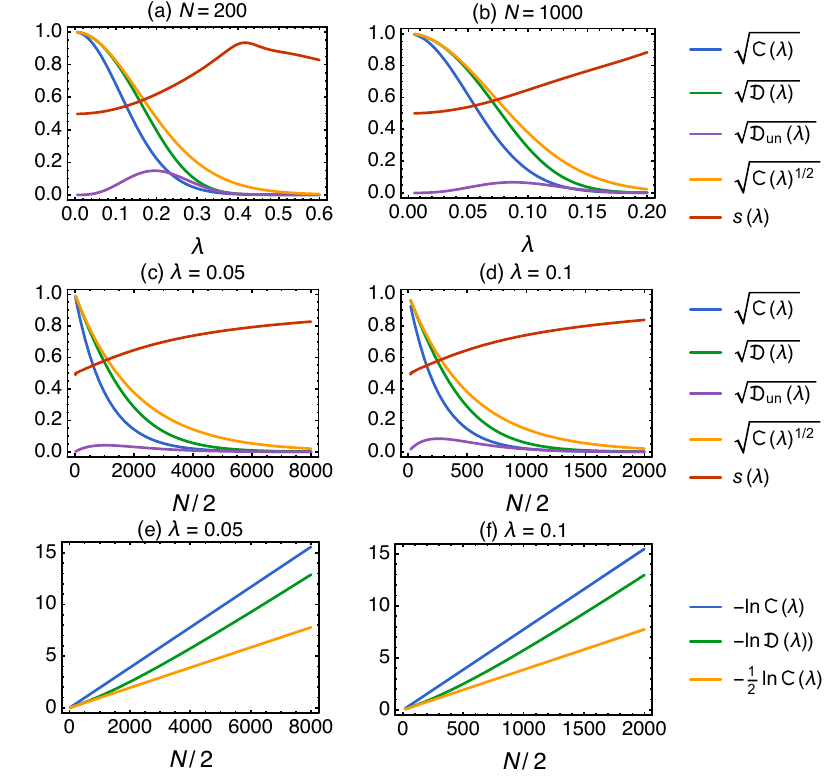}
\caption{
(Color online)
Various quantities,
$\sqrt{\mathcal{C}(\lambda)}$ (\ref{eq: define GOC}), $\sqrt{\mathcal{D}(\lambda)}$ (\ref{eq: define overlap of orthogonal components}), 
$\sqrt{\mathcal{D}^{\,}_{\mathrm{un}}(\lambda)}$ (\ref{eq: define un normalized D overlap}),
$\sqrt{\mathcal{C}(\lambda)^{1/2}}$, and 
$s(\lambda)$ (\ref{eq: sD postulate form})
for the driven Rice-Mele model (\ref{eq: driven RM model})  
are plotted as a function of $\lambda$ or $N$.
\label{Fig: CDD as a function of lambda and N}
         } 
\end{center}
\end{figure}

To further investigate the behavior of the normalized overlap $\sqrt{\mathcal{D}(\lambda)}$,
we compare it with the ground state overlap $\sqrt{\mathcal{C}(\lambda)}$ in Fig.\ \ref{Fig: CDD as a function of lambda and N}.
Notably, 
both $\sqrt{\mathcal{D}(\lambda)}$ [green curve] and $\sqrt{\mathcal{C}(\lambda)}$ [blue curve] 
decay monotonically as $N$ and $\lambda$ increase.
Moreover, $\sqrt{\mathcal{D}(\lambda)}$ exhibits a slower decay compared to $\sqrt{\mathcal{C}(\lambda)}$.
For further comparison, the unnormalized overlap $\sqrt{\mathcal{D}^{\,}_{\mathrm{un}}(\lambda)}$ [purple curve] 
is also depicted in Fig.\ \ref{Fig: CDD as a function of lambda and N}.

Given the explicit form of $\sqrt{\mathcal{D}^{\,}_{\mathrm{un}}(\lambda)}$ (\ref{eq: sDun vis final two level and k ind a}),
we may substitute it into Eq.\ (\ref{eq: inequality perp basis}) and apply the inequality of quantum speed limit (\ref{eq: bound from QSL})
to obtain the following bound on $\left|\mathcal{F}(\lambda)-\mathcal{C}(\lambda)\right|$
\begin{subequations}
\label{eq: new improved bound with alpha}
\begin{align}
&
\left|
\mathcal{F}(\lambda)
-
\mathcal{C}(\lambda)
\right|
\quad\leq\quad
g(\lambda),
\quad
g(\lambda)\:=
g^{\,}_{1}(\lambda)
+
g^{\,}_{2}(\lambda),
\label{eq: new improved bound with alpha a}
\\
&
g^{\,}_{1}(\lambda)\:=
\sin^2\widetilde{\mathcal{R}}(\lambda)
\mathcal{C}(\lambda)
\left|
-1
+
\alpha(\lambda)^{2}
\right|,
\label{eq: new improved bound with alpha b}
\\
&
g^{\,}_{2}(\lambda)\:=
\sin\Big(2\widetilde{\widetilde{\mathcal{R}}}(\lambda)\Big)\,
\mathcal{C}(\lambda)
\alpha(\lambda),
\label{eq: new improved bound with alpha c}
\end{align}
\end{subequations}
where $\widetilde{\mathcal{R}}(\lambda)$ is defined in Eq.\ (\ref{eq: bound from QSL})
and $\widetilde{\widetilde{\mathcal{R}}}(\lambda)$ is defined as
\begin{align}
&\widetilde{\widetilde{\mathcal{R}}}(\lambda)\:=
\min\left(\mathcal{R}(\lambda),\frac{\pi}{4}\right).
\label{eq: define R tilde tilde}
\end{align}
Note that combing the inequality (\ref{eq: new improved bound with alpha}) with the defining range of $\mathcal{F}(\lambda),$ i.e., 
$\mathcal{F}(\lambda)\in[0,1]$, yields
the following two-sided bound on the adiabatic fidelity $\mathcal{F}(\lambda)$
\begin{align}
\max\left(\mathcal{C}(\lambda)-g(\lambda),0\right)
\quad\leq\quad
\mathcal{F}(\lambda)
\quad\leq\quad
\min\left(\mathcal{C}(\lambda)+g(\lambda),1\right),
\nonumber
\end{align}
which provides a way to estimate the adiabatic fidelity $\mathcal{F}(\lambda)$ in terms of 
the ground state overlap $\mathcal{C}(\lambda)$ (\ref{eq: define GOC}) and the function $\mathcal{R}(\lambda)$ (\ref{eq: bound from QSL c}).

For comparison, let us revisit the inequality given by Eq.\ (\ref{eq: define an auxiliary function}), which is derived from Eq.\ (\ref{eq: inequality perp basis}) by substituting the overlap $\sqrt{\mathcal{D}^{\,}_{\mathrm{un}}(\lambda)}$ with its universal upper bound [see Eq.\ (\ref{eq: trivial upper bound for fidelity})]. 
When the quantum speed limit inequality from Eq.\ (\ref{eq: bound from QSL}) is applied to bound the Bures angle $\theta(\lambda)$ 
in Eq.\ (\ref{eq: define an auxiliary function}),
the following inequality was derived in Ref.\ \cite{Chen21}:
\begin{subequations}
\label{eq: summarized inequalities one side}
\begin{align}
&\left|\mathcal{F}(\lambda)-\mathcal{C}(\lambda)\right|
\;\leq\;
f(\lambda),
\quad
f(\lambda)\:= f^{\,}_{1}(\lambda)+ f^{\,}_{2}(\lambda),
\label{eq: summarized inequalities one side a}
\\
&f^{\,}_{1}(\lambda)\:=\sin^2\widetilde{\mathcal{R}}(\lambda)
\left|
1-
2\mathcal{C}(\lambda)
\right|,
\label{eq: summarized inequalities one side b}
\\
&f^{\,}_{2}(\lambda)\:=
\sin(2\widetilde{\widetilde{\mathcal{R}}}(\lambda))
\sqrt{\mathcal{C}(\lambda)}\sqrt{1-\mathcal{C}(\lambda)},
\label{eq: summarized inequalities one side c}
\end{align}
where $\widetilde{\mathcal{R}}(\lambda)$ and $\widetilde{\widetilde{\mathcal{R}}}(\lambda)$
are defined in Eq.\ (\ref{eq: bound from QSL}) and Eq.\ (\ref{eq: define R tilde tilde}), respectively.
\end{subequations}

\begin{figure}[t]
\begin{center}
\includegraphics[width=0.6\textwidth]{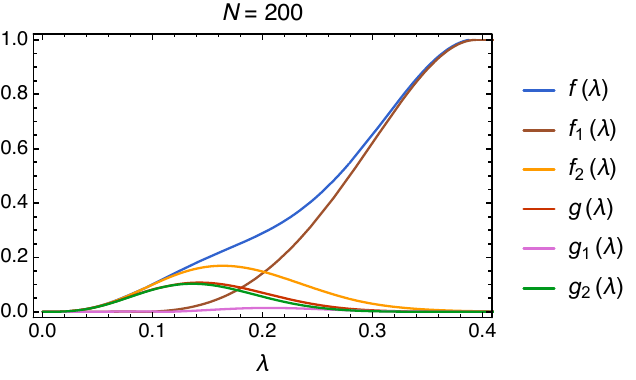}
\caption{
(Color online)
Compare the behavior of the function
$g(\lambda)=g^{\,}_{1}(\lambda)+g^{\,}_{2}(\lambda)$ (\ref{eq: new improved bound with alpha})
with the function
$f(\lambda)=f^{\,}_{1}(\lambda)+f^{\,}_{2}(\lambda)$ (\ref{eq: summarized inequalities one side})
for $N=200.$
\label{Fig: compare different bounds fg}
         }
\end{center}
\end{figure}

In Fig.\ \ref{Fig: compare different bounds fg}, we compare the two upper bounds on $|\mathcal{F}(\lambda)-\mathcal{C}(\lambda)|$:
$g(\lambda)=g^{\,}_{1}(\lambda)+g^{\,}_{2}(\lambda)$ 
[Eq.\ (\ref{eq: new improved bound with alpha})] and $f(\lambda)=f^{\,}_{1}(\lambda)+f^{\,}_{2}(\lambda)$ [Eq.\ (\ref{eq: summarized inequalities one side})].
The former one corresponds to the case where the unnormalized overlap 
$\sqrt{\mathcal{D}^{\,}_{\mathrm{un}}(\lambda)}$ 
takes the explicit form given in Eq.\ (\ref{eq: sDun vis final two level and k ind a}),
whereas the latter one is obtained by substituting $\sqrt{\mathcal{D}^{\,}_{\mathrm{un}}(\lambda)}$ 
with its universal upper bound [Eq.\ (\ref{eq: trivial upper bound for fidelity})].
The result shows that the function $f(\lambda)$ [blue curve] 
increases monotonically with $\lambda$,
whereas the function $g(\lambda)$ [red curve] exhibits a bell-shaped profile.
This indicates that $g(\lambda)$ serves as a better upper bound compared to $f(\lambda)$.
The enhanced performance of $g(\lambda)$ at larger $\lambda$ values stems from the function $g^{\,}_{1}(\lambda)$ [Eq.\ (\ref{eq: new improved bound with alpha b})], 
where the exponentially decaying factor  $\mathcal{C}(\lambda)$ is extracted, offering a contrast to $f^{\,}_{1}(\lambda)$ 
[Eq.\ (\ref{eq: summarized inequalities one side b})].
Meanwhile, the difference between the function $g^{\,}_{2}(\lambda)$ [green curve] and
the function $f^{\,}_{2}(\lambda)$ [orange curve] is not significant.
Consequently,
among the two inequalities, Eq.\ (\ref{eq: new improved bound with alpha}) and Eq.\ (\ref{eq: summarized inequalities one side}),
the former provides a better estimate for the adiabatic fidelity $\mathcal{F}(\lambda).$

\begin{figure}[t]
\begin{center}
\includegraphics[width=0.45\textwidth]{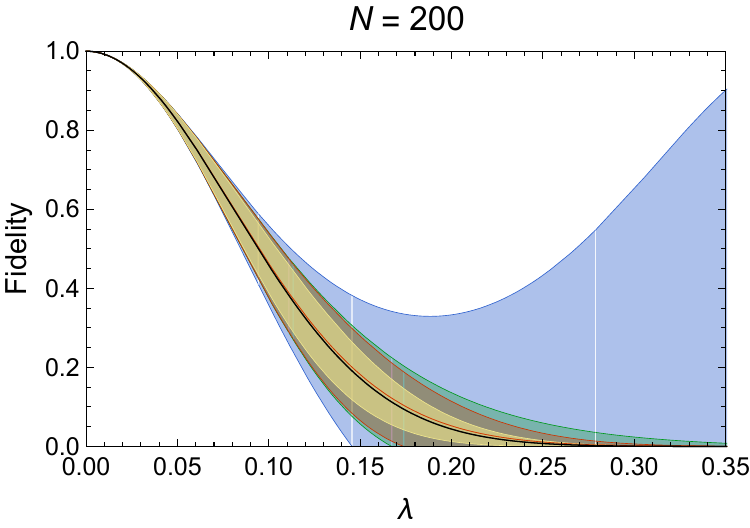}
\includegraphics[width=0.45\textwidth]{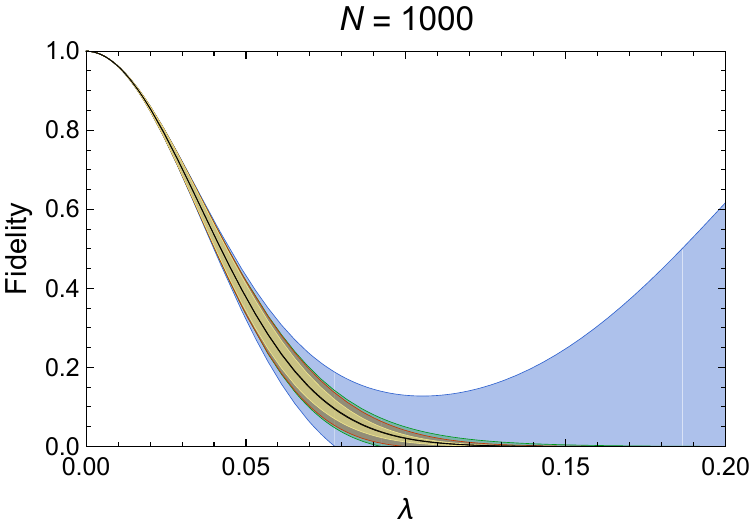}
\caption{
(Color online)
Bounds on the adiabatic fidelity $\mathcal{F}(\lambda)$ for $N=200$ and $N=1000$
using Eq.\ (\ref{eq: summarized inequalities one side}) [blue-shaded region], 
Eq.\ (\ref{eq: new improved bound with alpha}) [red-shaded region],
and Eq.\ (\ref{eq: a new improved bound}) with $s=1/2$ [green-shaded region] and $s=1$ [yellow-shaded region].
For both figures, the actual adiabatic fidelity $\mathcal{F}(\lambda)$ is the black curve,
while the red curve is for the ground state overlap $\mathcal{C}(\lambda)$,
which is, however, not distinct from $\mathcal{F}(\lambda)$.
\label{Fig: improved bounds}
         }
\end{center}
\end{figure}

In Fig.\ \ref{Fig: improved bounds}, we offer a comparison: estimates of the adiabatic fidelity $\mathcal{F}(\lambda)$ derived from Eq.\ (\ref{eq: summarized inequalities one side}) are represented by a blue-shaded region, while those from the improved inequality, Eq.\ (\ref{eq: new improved bound with alpha}), 
appear in a red-shaded region.
The improvement in estimation achieved using the improved inequality is evident.
Specifically, the improved estimate derived from Eq.\ (\ref{eq: new improved bound with alpha}) 
is effective even for a system size of $N=\mathcal{O}(10^2)$, 
which is the same large $N$ limit beyond which the ground state overlap $\mathcal{C}(\lambda)$ 
can be accurately approximated by a form of generalized orthogonality catastrophe [see Eq.\ (\ref{eq: scaling form of GOC})].
It is noteworthy that previous estimates on the adiabatic fidelity $\mathcal{F}(\lambda)$ obtained by Ref.\ \cite{Lychkovskiy17} and Ref.\ \cite{Chen21}  
were only effectively applicable for larger system sizes, specifically $N=\mathcal{O}(10^4)$ and $N=\mathcal{O}(10^3),$ respectively.

\section{Asymptotic form of the overlap $\mathcal{D}(\lambda)$ and implications}
\label{sec: Improved bounds on adiabatic fidelity}


While the function 
$g(\lambda)$ [Eq.\ (\ref{eq: new improved bound with alpha})] offers a better upper bound on 
$|\mathcal{F}(\lambda)-\mathcal{C}(\lambda)|$ than 
the function $f(\lambda)$ [Eq.\ (\ref{eq: summarized inequalities one side})], determining the overlap 
$\mathcal{D}(\lambda)$ explicitly can be challenging for generic many-body systems. 
An explicit form of the overlap $\mathcal{D}(\lambda)$ is crucial for the enhancement in $g(\lambda)$.
We thus seek for a universal scaling form of $\mathcal{D}(\lambda)$,
upon which an estimate of upper bound on $|\mathcal{F}(\lambda)-\mathcal{C}(\lambda)|$ can be obtained by means of Eq.\ (\ref{eq: inequality perp basis})
without calculating $\mathcal{D}(\lambda)$ from scratch.
In light of the reasoning of almost-orthogonality presented in Sec.\ \ref{sec: Perspective from almost orthogonality},
we consider the following ratio
\begin{align}
s(\lambda)\:=\frac{\ln\mathcal{D}(\lambda)}{\ln\mathcal{C}(\lambda)}.
\label{eq: sD postulate form}
\end{align}
The value of $s(\lambda)\geq0$ indicates how fast the overlap $\mathcal{D}(\lambda)$ decays compared to the overlap $\mathcal{C}(\lambda).$
If the ratio $s(\lambda)$ takes values in $[0,1]$, then the overlap $\mathcal{D}(\lambda)$ decays not faster than the overlap $\mathcal{C}(\lambda)$ does.

\begin{figure}[t]
\begin{center}
\includegraphics[width=0.6\textwidth]{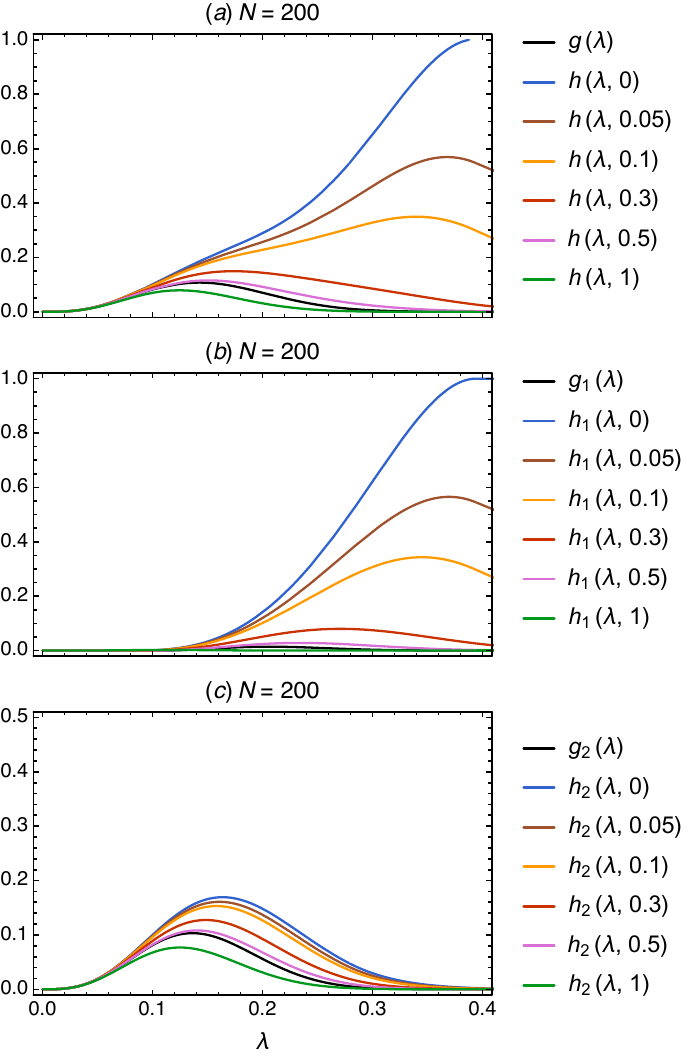}
\caption{
(Color online)
Compare the behavior of the function
$h^{\,}_{}(\lambda,s)$ (\ref{eq: a new improved bound a}),
$h^{\,}_{1}(\lambda,s)$ (\ref{eq: a new improved bound b}),
and
$h^{\,}_{2}(\lambda,s)$ (\ref{eq: a new improved bound c})
with different constant values of $s.$
As for a comparison, the function $g(\lambda)=g^{\,}_{1}(\lambda)+g^{\,}_{2}(\lambda)$ (\ref{eq: new improved bound with alpha})
is also depicted.
\label{Fig: compare different bounds h only}
         }
\end{center}
\end{figure}

By substituting $\mathcal{D}(\lambda)$ from Eq.\ (\ref{eq: sD postulate form}), which is given by 
$\mathcal{D}(\lambda) = \left(\mathcal{C}(\lambda)\right)^{s}$, 
into Eq.\ (\ref{eq: inequality perp basis}) 
and then applying the quantum speed limit inequality from (\ref{eq: bound from QSL}), 
we obtain the following inequality:
\begin{subequations}
\label{eq: a new improved bound}
\begin{align}
&
\left|
\mathcal{F}(\lambda)
-
\mathcal{C}(\lambda)
\right|
\quad\leq\quad
h(\lambda,s)\equiv
h^{\,}_{1}(\lambda,s)
+
h^{\,}_{2}(\lambda,s),
\label{eq: a new improved bound a}\\
&
h^{\,}_{1}(\lambda,s)\:=
\sin^2\widetilde{\mathcal{R}}(\lambda)
\mathcal{C}(\lambda)
\left|
1
-
\mathcal{C}(\lambda)^{s-1}
+
\mathcal{C}(\lambda)^{s}
\right|,
\label{eq: a new improved bound b}\\
&
h^{\,}_{2}(\lambda,s)\:=
\sin\Big(2\widetilde{\widetilde{\mathcal{R}}}(\lambda)\Big)
\sqrt{\mathcal{C}(\lambda)^{s+1}}
\sqrt{1-\mathcal{C}(\lambda)},
\label{eq: a new improved bound c}
\end{align}
\end{subequations}
where $\widetilde{\mathcal{R}}(\lambda)$ and $\widetilde{\widetilde{\mathcal{R}}}(\lambda)$ are defined in Eqs.\ (\ref{eq: bound from QSL a}) 
and (\ref{eq: define R tilde tilde}), respectively.
The inequality (\ref{eq: a new improved bound}) should be compared with that of 
Eq.\ (\ref{eq: inequality when exact orthogonality holds final result}), 
Eq.\ (\ref{eq: new improved bound with alpha}),
and Eq.\ (\ref{eq: summarized inequalities one side}).
It is worth noting that when $s=0$, the function $h(\lambda,s=0)$ simplifies to the function $f(\lambda)$ in Eq.\ (\ref{eq: summarized inequalities one side}). 
Observe that the function $h(\lambda,s)$ from Eq.\ (\ref{eq: a new improved bound}) is expressed as
$
\sin\mathcal{R}(\lambda)\mathcal{C}(\lambda)\times(\cdots)$,
where the terms inside the parenthesis 
scale at most polynomially in $N$ and $\lambda$ provided $s$ is not too small.
The dominant factor in the function $h(\lambda,s)$ for large $N$ and large $\lambda$ is the 
$\sin\mathcal{R}(\lambda)\mathcal{C}(\lambda)$ component. 
This dominance is also observed in Eq.\ (\ref{eq: inequality when exact orthogonality holds final result}) 
and in the function $g(\lambda)$ from Eq.\ (\ref{eq: new improved bound with alpha}). 
However, this is not the case for the function $f(\lambda)$ in Eq.\ (\ref{eq: summarized inequalities one side}). 
In essence, only the behavior of the function $h^{\,}_{1}(\lambda,s)$ at large $\lambda$ values determines whether $h(\lambda,s)$ can serve as a suitable upper bound.

Although the value of $s$ in Eq.\ (\ref{eq: a new improved bound}) typically depends on both $\lambda$ and $N$, 
it is possible to approximate it using specific constant values, thereby facilitating the use of Eq.\ (\ref{eq: a new improved bound}) in the estimation of adiabatic fidelity $\mathcal{F}(\lambda)$. 
To illustrate this, we shall revisit the driven Rice-Mele model presented in Sec.\ \ref{sec: Example: non-interacting Hamiltonians}.
In Fig.\ \ref{Fig: compare different bounds h only},
we plot $h^{\,}_{}(\lambda,s)=h^{\,}_{1}(\lambda,s)+h^{\,}_{2}(\lambda,s)$
(\ref{eq: a new improved bound}) as a function of $\lambda$ for different constant values of $s=0, 0.05, 0.1, 0.3, 0.5, 1.$
We observe that as long as the value of $s$ is not too small, say, $s\gtrsim 0.3$, 
the function $h^{\,}_{1}(\lambda,s)$ [see Fig.\ \ref{Fig: compare different bounds h only}(b)] decays quickly at large $\lambda$,
which improves the tail behavior of the function $h^{\,}_{}(\lambda,s)$ [see Fig.\ \ref{Fig: compare different bounds h only}(a)].
On the other hand, the behavior of $h^{\,}_{2}(\lambda,s)$ [see Fig.\ \ref{Fig: compare different bounds h only}(c)] does not change significantly as 
the value of $s$ varies.
Nevertheless, as long as the value of $s$ is sufficiently large, the function $h^{\,}_{}(\lambda,s)$ is dominated by $h^{\,}_{2}(\lambda,s)$
and serves as a strong upper bound on $|\mathcal{F}(\lambda)-\mathcal{C}(\lambda)|$.
In Fig.\ \ref{Fig: improved bounds}, we plot 
bounds on the adiabatic fidelity $\mathcal{F}(\lambda)$ for $N=200$ and $N=1000$
using Eq.\ (\ref{eq: a new improved bound}) with $s=1/2$ [green-shaded region] and $s=1$ [yellow-shaded region].
Note that the case of $s=0$ reduces to the inequality (\ref{eq: summarized inequalities one side}) [blue-shaded region].
One observes that the difference between the green-shaded area and the red-shaded area is not significant.
That is to say, the result of taking $s=1/2$, namely, taking $\mathcal{D}(\lambda)\approx\sqrt{\mathcal{C}(\lambda)}$, is very close to the result obtained from the 
inequality (\ref{eq: new improved bound with alpha}), which is derived using an explicit form of the normalized overlap $\mathcal{D}(\lambda)$ 
(\ref{eq: D overlap analytical form}).
Consequently,
we also plot $\sqrt{\mathcal{C}(\lambda)}$ and $s(\lambda)$ (\ref{eq: sD postulate form}) in Fig.\ \ref{Fig: CDD as a function of lambda and N},
which shows that $\sqrt{\mathcal{C}(\lambda)}$ is slightly larger than $\mathcal{D}(\lambda)$.
Nevertheless, for the purpose of estimating the adiabatic fidelity $\mathcal{F}(\lambda)$ using the inequality (\ref{eq: a new improved bound}), 
replacing $s(\lambda)$ by a constant value (such as $1/2$)
may be a good approximation.

\begin{figure}[t]
\begin{center}
\includegraphics[width=0.4\textwidth]{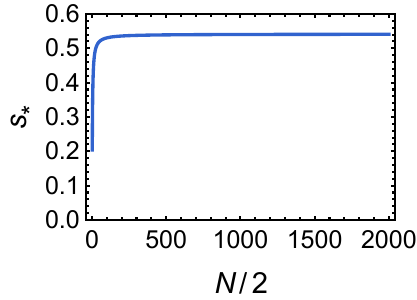}
\caption{
(Color online)
The ratio $s^{\,}_{*}$, defined in Eq.\ (\ref{eq: sD postulate form}) with $\lambda=\lambda^{\,}_{*}$ (so that $\mathcal{C}(\lambda^{\,}_{*})=1/e$),
is calculated numerically for the driven Rice-Mele model (\ref{eq: driven RM model})  
with the value of parameters shown in Eq.\ (\ref{eq: numerical parameter}).
Asymptotically, $s^{\,}_{*}=\mathcal{O}(1)$ as $N\to\infty.$
\label{Fig: Sstar}
         }
\end{center}
\end{figure}

\begin{figure}[t]
\begin{center}
\includegraphics[width=0.24\textwidth]{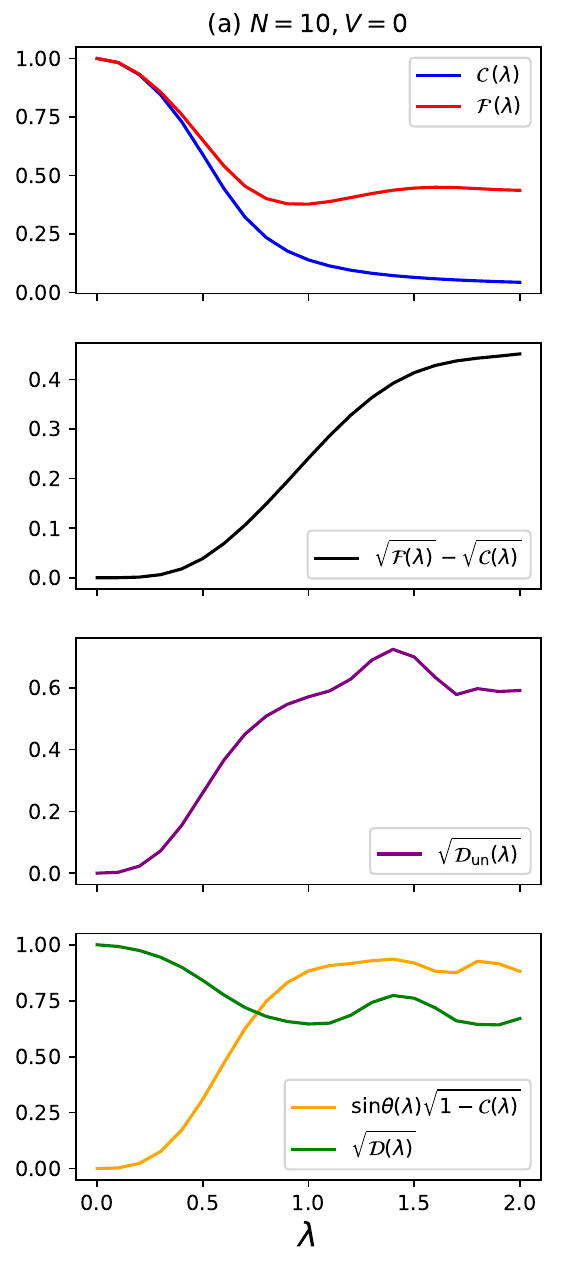}
\includegraphics[width=0.24\textwidth]{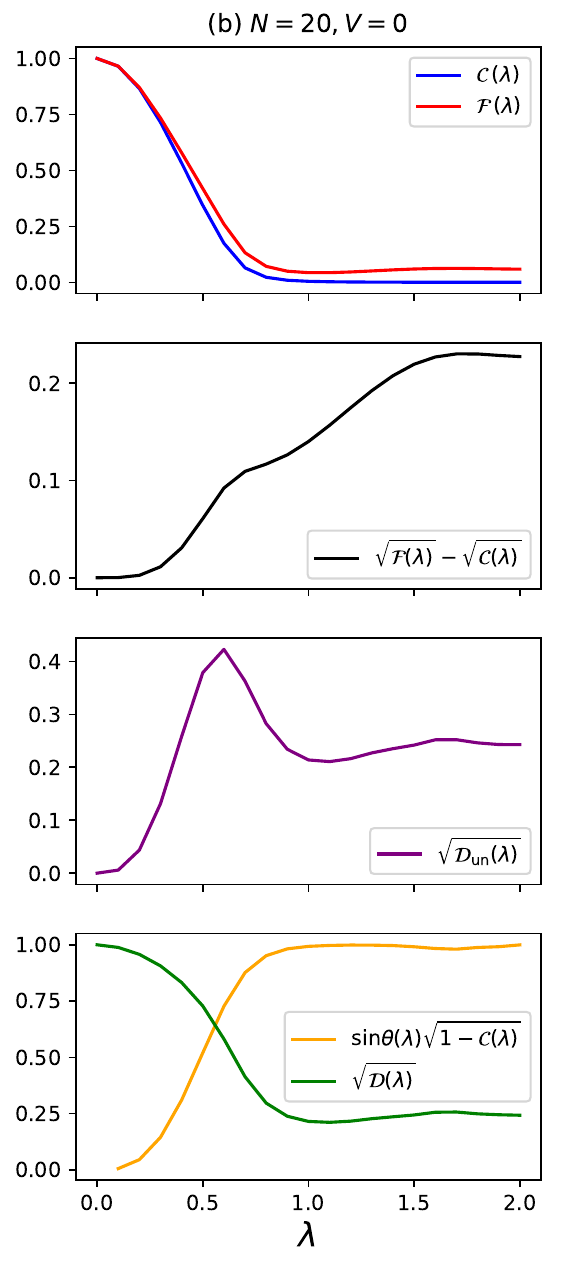}
\includegraphics[width=0.24\textwidth]{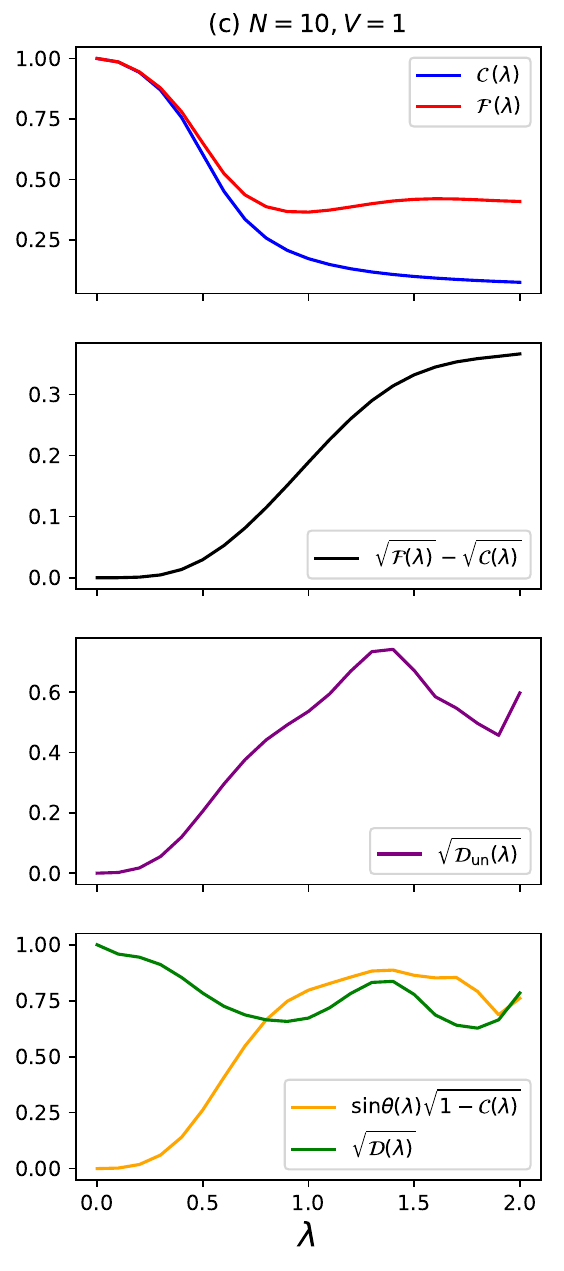}
\includegraphics[width=0.24\textwidth]{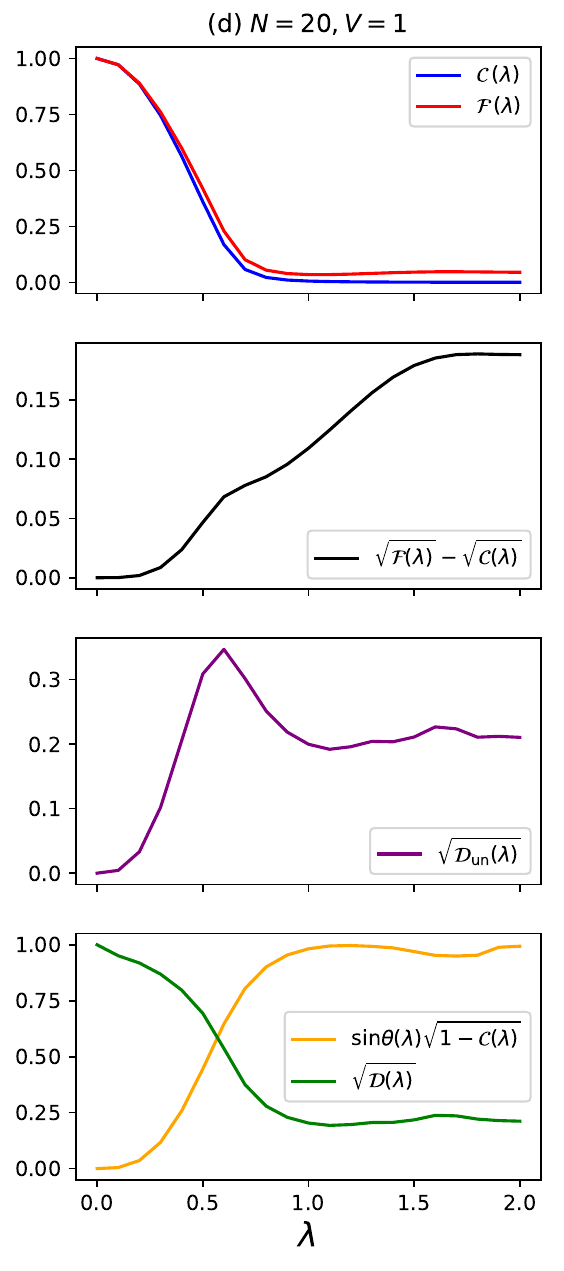}
\caption{
(Color online)
Various quantities are calculated numerically for the driven interacting Kitaev chain model (\ref{eq: driven Kitaev model})  
with the value of parameters shown in Eq.\ (\ref{eq: numerical parameter Kitaev})
for system size $N=10$ and $N=20$ with interaction strength $V=0$ and $V=1.$ 
Further explanation is provided in the main text.
\label{Fig: F minus C Kitaev}
         }
\end{center}
\end{figure}

\begin{figure}[t]
\begin{center}
\includegraphics[width=0.3\textwidth]{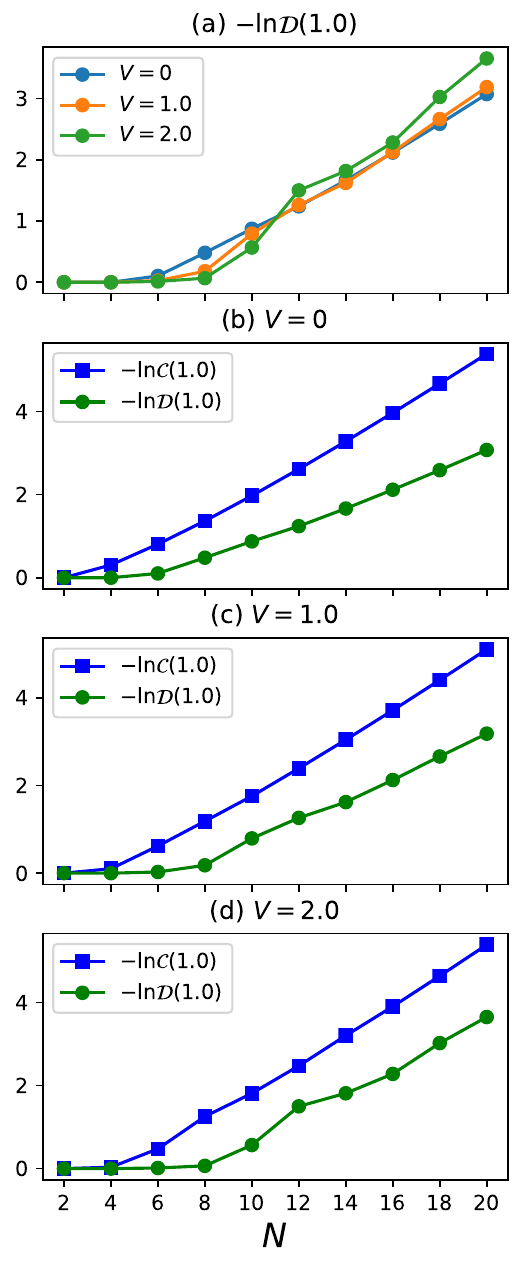}
\includegraphics[width=0.3\textwidth]{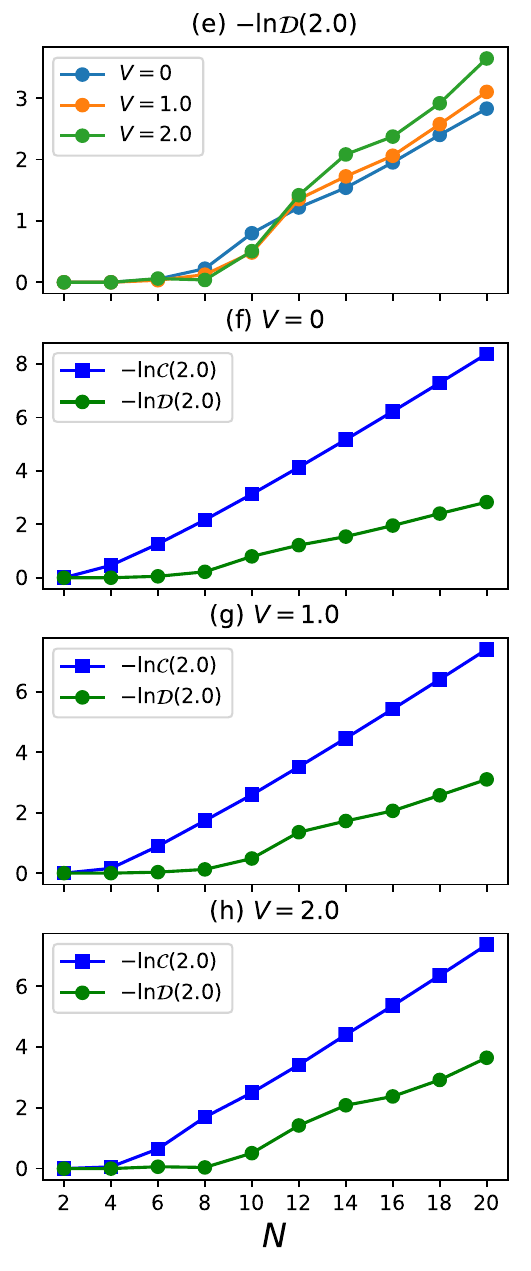}
\caption{
(Color online)
The quantities
$-\ln\mathcal{C}(\lambda)$ (\ref{eq: define GOC}) and $-\ln\mathcal{D}(\lambda)$ (\ref{eq: define overlap of orthogonal components})
for the driven interacting Kitaev model (\ref{eq: driven Kitaev model}) 
are plotted as a function of $N$ for different values of $\lambda$ ($\lambda=1$ in the left column and $\lambda=2$ in the right column) and $V=0,1,2$.
\label{Fig: CDD as a function of lambda and N Kitaev}
         } 
\end{center}
\end{figure}

\subsection*{Condition for adiabaticity breakdown revisited}

Before concluding this section, 
let us discuss implications of a condition for adiabaticity breakdown 
using inequality (\ref{eq: a more useful form for triangle inequality}). 
Recall that a large class of driven many-body systems exists for which the condition    
\begin{align}
\delta V^{\,}_{N}/C^{\,}_{N}=0
\quad
\text{as}
\quad
N\to\infty.
\end{align}
holds \cite{Lychkovskiy17}.
Here, $\delta V^{\,}_{N}$ is introduced in Eq.\ (\ref{eq: bound from QSL special form}), and 
$C^{\,}_{N}$ refers to the exponent given in Eq.\ (\ref{eq: scaling form of GOC}).
Define $\lambda^{\,}_{*}$ as an {\it adiabatic mean free path}
so that $\mathcal{F}(\lambda)\geq e^{-1}$ for $\lambda\leq\lambda^{\,}_{*}.$
Here, $\lambda^{\,}_{*}$ is determined 
by $\mathcal{C}(\lambda^{\,}_{*})=1/e$.
The relation $\mathcal{R}(\lambda^{\,}_{*})=\delta V^{\,}_{N}/(2\Gamma C^{\,}_{N})$ then 
follows from Eq.\ (\ref{eq: bound from QSL special form}).
It was shown in Ref.\ \cite{Lychkovskiy17} that,
in order to avoid adiabaticity breakdown, 
the driving rate $\Gamma$ of driven many-body systems must scale down with increasing system size $N$.
Using inequality (\ref{eq: a more useful form for  triangle inequality}),
we find
that quantum adiabaticity is maintained if $\Gamma\leq\Gamma^{\,}_{N}$ with
[see Appendix \ref{App3: Detail calculation for implication on adiabaticity breakdown} for a derivation] 
\begin{align}
&\Gamma^{\,}_{N}
\:=
\frac{1}{2}
\frac{\delta V^{\,}_{N}}{C^{\,}_{N}}
\frac{1}{\left(1-\epsilon-e^{-1/2}\right)^{2}}
M(s^{\,}_{*}),
\label{eq: scaling down of Gamma general form}
\end{align}
where
$\epsilon\in[0,1]$,
$M(s^{\,}_{*})\:=\sqrt{1-e^{-1}}\,e^{-s^{\,}_{*}/2}$ and $s^{\,}_{*}\:=s(\lambda^{\,}_{*})=-\ln\mathcal{D}(\lambda^{\,}_{*})$.
Consequently, the asymptotic form of $M(s^{\,}_{*})$ as $N\to\infty$ contributes to a multiplicative modification to the scaling form
$\Gamma^{\,}_{N}\sim\delta V^{\,}_{N}/C^{\,}_{N}$ found previously in Refs.\ \cite{Lychkovskiy17,Chen21}.
Two key observations can be made:
(i)
If $s^{\,}_{*}=\mathcal{O}(1)$ as $N\to\infty$, then $M(s^{\,}_{*})=\mathcal{O}(1)$.
(ii)
For 
$s^{\,}_{*}=\mathcal{O}(N^{c})$ with $c$ being a real number as $N\to\infty,$ then $M(s^{\,}_{*})=\mathcal{O}(e^{-N^{c}}).$
Further, it is verified that when the leading asymptotics of 
$-\ln\mathcal{D}(\lambda)$ is 
proportional to that of
$-\ln\mathcal{C}(\lambda)$ as $N\to\infty,$
then the condition $s^{\,}_{*}=\mathcal{O}(1)$ is satisfied.
Specifically, for the driven Rice-Mele model (\ref{eq: driven RM model}), we determine that 
$s^{\,}_{*}=\mathcal{O}(1)$ as $N\to\infty$ [refer to Fig.\ \ref{Fig: Sstar}], which implies
$M(s^{\,}_{*})=\mathcal{O}(1)$.

\section{Illustrative example II: interacting fermions}
\label{sec: Example: interacting fermions}

Up to this point, our general results from Secs.\ \ref{sec: a motivating limit} and \ref{sec: Reverse triangle inequalities},
have been illustrated using the driven Rice-Mele model (\ref{eq: driven RM model}), 
a model of quadratic fermions 
characterized by the unique property where the underlying Hilbert space is constructed as a direct product of single-particle states. 
A pivotal question emerges regarding whether the  phenomenon of almost-orthogonality between vectors in the complement of the initial state exists in typical many-body systems, especially those governed by nonintegrable interacting Hamiltonians.
In this section, we respond to the question affirmatively.
To demonstrate this, we analyze an interacting Kitaev chain model defined by the following Hamiltonian
\begin{align}
H^{\,}_{\mathrm{K}}
\:=&
\sum^{N}_{j=1}
\left[
\left(
-J
c^{\dag}_{j}c^{\,}_{j+1}
+
\Delta
c^{\dag}_{j}c^{\dag}_{j+1}
+
\mathrm{h.c.}
\right)
+
Vn^{\,}_{j}n^{\,}_{j+1}
\right]
+
\sum^{N}_{j=1}
\mu(\lambda)n^{\,}_{j},
\label{eq: driven Kitaev model}
\end{align}
where $n^{\,}_{j}\:=c^{\dag}_{j}c^{\,}_{j}$
is the number operator of fermions at lattice site $j$, $N$ the number of lattice sites,
$J$ the hopping amplitude, $\Delta$ the superconducting pairing amplitude, 
$V$ the strength of nearest-neighbor Coulomb repulsion, 
and $\mu(\lambda)=\mu^{\,}_{0}\lambda$ (with $\mu^{\,}_{0}\in\mathbb{R}$) the time-dependent chemical potential. 
If $V=0,$ the Hamiltonian (\ref{eq: driven Kitaev model}) reduces to that of the Kitaev model of one-dimensional $p$-wave superconductors \cite{Kitaev00}.
We shall consider the Hamiltonian (\ref{eq: driven Kitaev model}) with periodic boundary conditions and the sector of odd fermion parity.
To ensure that our numerical simulation results reflect generic features, 
we avoid selecting parameter values corresponding to solvable points \cite{Chepiga23}.
For concreteness, 
we choose the following parameter values
\begin{align}
(J,\Delta, \mu^{\,}_{0}, \Gamma) = (1.0,0.8,3.0,1.0).
\label{eq: numerical parameter Kitaev}
\end{align}

In Fig.\ \ref{Fig: F minus C Kitaev}, we plot various quantities as functions of
$\lambda$ for the driven interacting Kitaev model (\ref{eq: driven Kitaev model}).
Specifically, panels (a) and (c) depict results for a system size of $N=10$,
while panels (b) and (d) illustrate results for $N=20$.
The interaction strength $V=0$ is chosen for panels (a) and (b), and
$V=1$ for panels (c) and (d).
The behavior of curves in panel (a) is quantitatively similar to those of the driven Rice-Mele model with $N=10$ presented in Fig.\ \ref{Fig: F minus C}(a). 
For the increased system size $N=20$, which approaches the computational limits of the exact diagonalization method,
the notable characteristic is the expedited decay of 
$\mathcal{F}(\lambda)$,
$\mathcal{C}(\lambda),$ and $\mathcal{D}(\lambda),$ 
as illustrated in Fig.\ \ref{Fig: F minus C Kitaev}(b). 
This trend aligns with observations for the driven Rice-Mele model shown in Fig.\ \ref{Fig: F minus C}.
Panels (c) and (d) of Fig.\ \ref{Fig: F minus C Kitaev} represent the case of the driven {\it interacting} Kitaev model with $V=1$ for system sizes
$N=10$ and $N=20,$ respectively.
Compared with their non-interacting counterparts [namely, panels (a) and (b)], 
the differences between panels (c) and (a), as well as between panels (d) and (b), are not significant. 
Hence, we anticipate that the phenomenon of almost-orthogonality between vectors in the complement of the initial state should exist
even in the presence of interacting Hamiltonians.
To further elucidate this point, we plot in Fig.\ \ref{Fig: CDD as a function of lambda and N Kitaev} the quantities
$-\ln\mathcal{C}(\lambda)$ (\ref{eq: define GOC}) and $-\ln\mathcal{D}(\lambda)$ (\ref{eq: define overlap of orthogonal components})
for the driven interacting Kitaev model (\ref{eq: driven Kitaev model}) as a function of system size $N$ for
$\lambda=1, 2$ and interaction strength $V=0,1,2$. 
According to all the panels of Fig.\ \ref{Fig: CDD as a function of lambda and N Kitaev}, 
the normalized overlap $\mathcal{D}(\lambda)$ decays with increasing system size $N$, 
a characteristic of almost-orthogonality in complementary subspace.

As a final note, 
we would like to highlight two additional notable numerical findings from Fig.\ \ref{Fig: CDD as a function of lambda and N Kitaev}.
First, the ground state overlap $\mathcal{C}(\lambda)$ decays faster than the normalized overlap $\mathcal{D}(\lambda)$,
a feature found similarly in the driven Rice-Mele model (see Fig.\ \ref{Fig: CDD as a function of lambda and N}).
Second, the decay exponent of $\mathcal{D}(\lambda)$ increases with increasing interaction strength $V$.
This observation may suggest that the presence of interactions leads to a distribution of vectors in the underlying Hilbert space that appears more ``random'',
resulting in a more rapid decay of the overlap $\mathcal{D}(\lambda)$ as the interaction strength increases (for a fixed system size).
The extent to which these two observations are universal, however, warrants further investigation.

\section{Summary and Outlook}
\label{sec: Discussion and Conclusion}


This study elucidates the reasons behind the frequent observation in quantum many-body systems where the adiabatic fidelity, $\mathcal{F}(\lambda)$, 
and the overlap between the initial and instantaneous ground states, $\mathcal{C}(\lambda)$, exhibit nearly identical values in numerous instances, 
especially in regions with small evolution parameters or large system sizes.
While this observation can be rationalized in the region of small evolution parameter $\lambda$ using detailed perturbation theory, 
which shows that the difference between $\mathcal{F}(\lambda)$ and $\mathcal{C}(\lambda)$ only appears at the $\lambda^{3}$ order, 
a thorough explanation for the region of large system size fundamentally hinges on an intrinsic property of quantum many-body systems: {\it the almost-orthogonality of random vectors}. 
Specifically, this work details how the almost-orthogonality in the complementary space of the initial state,
as exhibited by an exponentially decaying normalized overlap $\mathcal{D}(\lambda)$ (\ref{eq: define overlap of orthogonal components}) 
and a small value of unnormalized overlap $\mathcal{D}^{\,}_{\mathrm{un}}(\lambda)$ (\ref{eq: define un normalized D overlap}), controls an upper bound 
on the difference between $\sqrt{\mathcal{F}(\lambda)}$ and $\sqrt{\mathcal{C}(\lambda)}$ [see Eq.\ (\ref{eq: a more useful form for triangle inequality})]. 
To support these general, model-independent findings, numerical studies were conducted on both a driven Rice-Mele model and a driven interacting Kitaev model.

As a secondary result, our study provides improved estimates for the adiabatic fidelity $\mathcal{F}(\lambda)$. 
These estimates rely on an explicit representation of the normalized overlap 
$\mathcal{D}(\lambda)$ as a specific function of the ground state overlap $\mathcal{C}(\lambda)$, 
leveraging the concept of almost-orthogonality.
Using the driven Rice-Mele model as an illustration, we demonstrated that 
these refined estimates perform well down to system sizes $N=\mathcal{O}(10^2)$—the same threshold beyond which the ground state 
overlap $\mathcal{C}(\lambda)$ is accurately captured by a generalized orthogonality–catastrophe form.
These results distinctly outperform the earlier estimates of Refs.\ \cite{Lychkovskiy17,Chen21}, 
which achieve good accuracy mainly for $N\gtrsim 10^3$, marking a notable advance in the precision and practical utility of adiabatic-fidelity estimates.

There are two key observations identified through our numerical analysis that warrant further attention:
(i) The ground state overlap $\mathcal{C}(\lambda)$ decays more rapidly than the normalized overlap $\mathcal{D}(\lambda)$,
and (ii) the decay exponent of $\mathcal{D}(\lambda)$ increases with increasing interaction strength.
Future work should investigate whether these two observations persist across a broader range of models and explore the potential connection between the degree of {\it almost-orthogonality} and the dichotomy between chaotic and integrable many-body systems.

We conclude by noting that it would be worthwhile to pursue a rigorous proof (similar to that in Ref.\ \cite{Gebert14}) to confirm the existence of subspace almost-orthogonality. 
Additionally, it would be intriguing to explore how subspace almost-orthogonality might affect other driven many-body systems known to exhibit orthogonality catastrophe in the full Hilbert space, 
such as those with time-dependent impurities \cite{Knap12} or undergoing quantum quenches \cite{Schiro14}.


\section*{Acknowledgements}
The numerical calculations of Sec.\ \ref{sec: Example: interacting fermions} were accomplished using QuSpin \cite{Weinberg17,Weinberg19}.


\paragraph{Funding information}
This work is part of the Adiabatic Protocols in Extended Quantum Systems project, 
funded by the Dutch Research Council (NWO) under Project No 680-91-130.
During the revision of this work,
J.-H.\ C. was supported by the National Science and Technology
Council (NSTC) of Taiwan under Grant No.\ 113-2112-M-008-037-MY3.

\begin{appendix}

\section{Alternative derivation of Eq.\ (\ref{eq: inequality perp basis})}
\label{App0: Perturbative derivation of inequality}


We rederive inequality (\ref{eq: inequality perp basis})—originally obtained in Ref.~\cite{Chen21}—using a more compact projection-operator technique.
First, we compute $\mathcal{F}(\lambda)$ using Eq.\ (\ref{eq: expansion of states}),
\begin{align}
\mathcal{F}(\lambda)
&=
|\langle\Phi^{\,}_{\lambda}|(P+Q)|\Psi^{\,}_{\lambda}\rangle|^2
\nonumber\\
&=
|\langle\Phi^{\,}_{\lambda}|P|\Psi^{\,}_{\lambda}\rangle+\langle\Phi^{\,}_{\lambda}|Q|\Psi^{\,}_{\lambda}\rangle|^2
\nonumber\\
&=
|\langle\Phi^{\,}_{\lambda}|P|\Psi^{\,}_{\lambda}\rangle|^2
+
|\langle\Phi^{\,}_{\lambda}|Q|\Psi^{\,}_{\lambda}\rangle|^2
+
2\Re
\Big(
\langle\Psi^{\,}_{\lambda}|P|\Phi^{\,}_{\lambda}\rangle\langle\Phi^{\,}_{\lambda}|Q|\Psi^{\,}_{\lambda}\rangle
\Big).
\label{eq: F adb in perp basis}
\end{align}

Combing (i) Eq.\ (\ref{eq: F adb in perp basis}),
(ii) the triangle inequality for absolute value, 
(iii) the inequality $\Re(z)\leq|z|$ for $z\in\mathbb{C}$,
and (iv) Eqs.\  (\ref{eq: properties by constructions}) and (\ref{eq: define overlap of orthogonal components})
yields the following chain of inequalities
\begin{align}
\left|
\mathcal{F}(\lambda)
-
\mathcal{C}(\lambda)
\right|
\stackrel{\rm (i)}{=}&\,
\Bigg|
|\langle\Phi^{\,}_{\lambda}|P|\Psi^{\,}_{\lambda}\rangle|^2
-
\mathcal{C}(\lambda)
+
|\langle\Phi^{\,}_{\lambda}|Q|\Psi^{\,}_{\lambda}\rangle|^2
+
2\Re
\Big(
\langle\Psi^{\,}_{\lambda}|P|\Phi^{\,}_{\lambda}\rangle\langle\Phi^{\,}_{\lambda}|Q|\Psi^{\,}_{\lambda}\rangle
\Big)
\Bigg|
\nonumber\\
\stackrel{\rm (ii)}{\leq}&\,
\Bigg|
|\langle\Phi^{\,}_{\lambda}|P|\Psi^{\,}_{\lambda}\rangle|^2
-
\mathcal{C}(\lambda)
+
|\langle\Phi^{\,}_{\lambda}|Q|\Psi^{\,}_{\lambda}\rangle|^2
\Bigg|
+
2
\Bigg|
\Re
\Big(
\langle\Psi^{\,}_{\lambda}|P|\Phi^{\,}_{\lambda}\rangle\langle\Phi^{\,}_{\lambda}|Q|\Psi^{\,}_{\lambda}\rangle
\Big)
\Bigg|
\nonumber\\
\stackrel{\rm (iii)}{\leq}&\,
\Bigg|
|\langle\Phi^{\,}_{\lambda}|P|\Psi^{\,}_{\lambda}\rangle|^2
-
\mathcal{C}(\lambda)
+
|\langle\Phi^{\,}_{\lambda}|Q|\Psi^{\,}_{\lambda}\rangle|^2
\Bigg|
+
2
\left|
\langle\Psi^{\,}_{\lambda}|P|\Phi^{\,}_{\lambda}\rangle
\right|
\left|
\langle\Phi^{\,}_{\lambda}|Q|\Psi^{\,}_{\lambda}\rangle
\right|
\nonumber\\
\stackrel{\rm (iv)}{=}&\,
\left|
-\sin^2\theta(\lambda)\mathcal{C}(\lambda)
+
\mathcal{D}^{\,}_{\mathrm{un}}(\lambda)
\right|
+
2
\cos\theta(\lambda)
\sqrt{\mathcal{C}(\lambda)}
\sqrt{\mathcal{D}^{\,}_{\mathrm{un}}(\lambda)}.
\label{eq: inequality perp basis app}
\end{align}

\subsection*{A further alternative derivation}

Alternatively, the inequality (\ref{eq: inequality perp basis}) 
can also be obtained using the set of triangle inequality (\ref{eq: reverse triangle inequalities}).
We begin with the upper bound on $\sqrt{\mathcal{F}(\lambda)}$ from Eq.\ (\ref{eq: reverse triangle inequalities 1}),
$
\sqrt{\mathcal{F}(\lambda)}
\leq
\cos^{\,}\theta(\lambda)\sqrt{\mathcal{C}(\lambda)}
+
\sqrt{\mathcal{D}^{\,}_{\mathrm{un}}(\lambda)}$,
and then calculate $\mathcal{F}(\lambda)-\mathcal{C}(\lambda)$,
\begin{align}
\mathcal{F}(\lambda)
-
\mathcal{C}(\lambda)
&
\quad\leq\quad
-\sin^{2}\theta(\lambda)\mathcal{C}(\lambda)
+
\mathcal{D}^{\,}_{\mathrm{un}}(\lambda)
+
2\cos^{\,}\theta(\lambda)\sqrt{\mathcal{C}(\lambda)}\sqrt{\mathcal{D}^{\,}_{\mathrm{un}}(\lambda)}
\nonumber\\
&
\quad\leq\quad
\left|
-\sin^{2}\theta(\lambda)\mathcal{C}(\lambda)
+
\mathcal{D}^{\,}_{\mathrm{un}}(\lambda)
\right|
+
2\cos^{\,}\theta(\lambda)\sqrt{\mathcal{C}(\lambda)}\sqrt{\mathcal{D}^{\,}_{\mathrm{un}}(\lambda)},
\label{eq: reverse triangle inequalities app part 1}
\end{align}
where the last step is obtained after using the inequality, $x+y\leq|x+y|$ for $x,y\in\mathbb{R}.$
Similarly, we consider the lower bound on $\sqrt{\mathcal{F}(\lambda)}$ from Eq.\ (\ref{eq: reverse triangle inequalities 2}),
$
\sqrt{\mathcal{F}(\lambda)}
\geq
\left|
\sqrt{\mathcal{D}^{\,}_{\mathrm{un}}(\lambda)}
-
\cos\theta(\lambda)\,\sqrt{\mathcal{C}(\lambda)}
\right|$,
and then calculate $\mathcal{F}(\lambda)-\mathcal{C}(\lambda)$,
\begin{align}
\mathcal{F}(\lambda)
-
\mathcal{C}(\lambda)
&
\quad\geq\quad
-\sin^{2}\theta(\lambda)\mathcal{C}(\lambda)
+
\mathcal{D}^{\,}_{\mathrm{un}}(\lambda)
-
2\cos^{\,}\theta(\lambda)\sqrt{\mathcal{C}(\lambda)}\sqrt{\mathcal{D}^{\,}_{\mathrm{un}}(\lambda)}
\nonumber\\
&
\quad\geq\quad
-
\left|
-\sin^{2}\theta(\lambda)\mathcal{C}(\lambda)
+
\mathcal{D}^{\,}_{\mathrm{un}}(\lambda)
\right|
-
2\cos^{\,}\theta(\lambda)\sqrt{\mathcal{C}(\lambda)}\sqrt{\mathcal{D}^{\,}_{\mathrm{un}}(\lambda)},
\label{eq: reverse triangle inequalities app part 2}
\end{align}
where the last step is obtained after using the inequality, $x+y\geq-|x+y|$ for $x,y\in\mathbb{R}.$
Combing Eqs.\ (\ref{eq: reverse triangle inequalities app part 1}) with (\ref{eq: reverse triangle inequalities app part 2}) 
yields the inequality (\ref{eq: inequality perp basis}).

\section{Perturbative expansion in $\lambda$}
\label{App1: Perturbative  expansion}

A detailed derivation of Eqs.\ (\ref{eq: perturbative AF and GOC}) and (\ref{eq: perturbative other quantities}) is provided.

\subsection{For instantaneous ground state}

We want to solve the instantaneous eigenvalue equation (\ref{eq: instantaneous eigenvalue equation}) perturbatively in $\lambda$,
\begin{align}
H^{\,}_{\lambda}
|\Phi^{\,}_{\lambda}\rangle
=
E^{\,}_{\mathrm{GS},\lambda}|\Phi^{\,}_{\lambda}\rangle,
\quad
H^{\,}_{\lambda}=H^{\,}_{0}+\lambda V,
\end{align}
given
$
H^{\,}_{0}
|\chi^{\,}_{n}\rangle
=
\varepsilon^{\,}_{n}|\chi^{\,}_{n}\rangle,
$
where $|\Phi^{\,}_{\lambda}\rangle$ is the ground state of $H^{\,}_{\lambda},$
and $\{|\chi^{\,}_{n}\rangle\}$ is a set of the complete orthonormal eigenstates of $H^{\,}_{0}$ with $|\chi^{\,}_{0}\rangle\equiv|\Psi^{\,}_{0}\rangle$ 
being the ground state of $H^{\,}_{0}$
and $n=0,1,\cdots$ labels distinct eigenstates.
Apply the standard Rayleigh-Schr\"odinger perturbation theory up to order $\lambda^2$
yields the following series,
\begin{align}
|\Phi^{\,}_{\lambda}\rangle
=&\,
\left(
1
-
\frac{\lambda^2}{2}
\sum^{\,}_{n\neq 0}
\frac{|V^{\,}_{n0}|^2}{(\varepsilon^{\,}_{0}-\varepsilon^{\,}_{n})^2}
\right)
|\chi^{\,}_{0}\rangle
+
\lambda
\sum^{\,}_{n\neq 0}
\frac{V^{\,}_{n0}}{\varepsilon^{\,}_{0}-\varepsilon^{\,}_{n}}\,
|\chi^{\,}_{n}\rangle
\nonumber\\
&+
\lambda^2
\sum^{\,}_{n\neq 0}
\frac{1}{\varepsilon^{\,}_{0}-\varepsilon^{\,}_{n}}
\left(
\sum^{\,}_{m\neq 0}
\frac{V^{\,}_{nm}V^{\,}_{m0}}{\varepsilon^{\,}_{0}-\varepsilon^{\,}_{m}}
-
\frac{V^{\,}_{00}V^{\,}_{n0}}{\varepsilon^{\,}_{0}-\varepsilon^{\,}_{n}}
\right)
|\chi^{\,}_{n}\rangle
+
\cdots,
\nonumber
\end{align}
where 
$
V^{\,}_{nm}\:=\langle\chi^{\,}_{n}|
V
|\chi^{\,}_{m}\rangle.
$
Hence, the following inner products are obtained,
\begin{subequations}
\label{eq: conclusion for instantaneous perturbation}
\begin{align}
\langle\chi^{\,}_{0}|\Phi^{\,}_{\lambda}\rangle=&\,
1
-
\frac{\lambda^2}{2}
\sum^{\,}_{n\neq 0}
\frac{|V^{\,}_{n0}|^2}{(\varepsilon^{\,}_{0}-\varepsilon^{\,}_{n})^2}
\nonumber\\
&+
\lambda^{3}
\Bigg(
\sum^{\,}_{n\neq 0}
\frac{V^{\,}_{00}|V^{\,}_{n0}|^2}{(\varepsilon^{\,}_{0}-\varepsilon^{\,}_{n})^{3}}
-
\sum^{\,}_{n\neq 0}
\frac{1}{(\varepsilon^{\,}_{0}-\varepsilon^{\,}_{n})^2}
\sum^{\,}_{m\neq 0}
\frac{\Re(V^{*}_{nm}V^{*}_{m0}V^{\,}_{n0})}{\varepsilon^{\,}_{0}-\varepsilon^{\,}_{m}}
\Bigg)+\cdots.
\label{eq: conclusion for instantaneous perturbation a}
\\
\langle\chi^{\,}_{n\neq0}|\Phi^{\,}_{\lambda}\rangle=&\,
\lambda
\frac{V^{\,}_{n0}}{\varepsilon^{\,}_{0}-\varepsilon^{\,}_{n}}
+
\lambda^2
\frac{1}{\varepsilon^{\,}_{0}-\varepsilon^{\,}_{n}}
\left(
\sum^{\,}_{m\neq 0}
\frac{V^{\,}_{nm}V^{\,}_{m0}}{\varepsilon^{\,}_{0}-\varepsilon^{\,}_{m}}
-
\frac{V^{\,}_{00}V^{\,}_{n0}}{\varepsilon^{\,}_{0}-\varepsilon^{\,}_{n}}
\right)+\cdots.
\label{eq: conclusion for instantaneous perturbation b}
\end{align}
\end{subequations}
Notice that both inner products, $\langle\chi^{\,}_{0}|\Phi^{\,}_{\lambda}\rangle$ and $\langle\chi^{\,}_{n\neq0}|\Phi^{\,}_{\lambda}\rangle$, are real-valued.

\subsection{For time-evolved state}

We want to solve the time-dependent Schr\"odinger equation (\ref{eq: schroedinger equation}) perturbatively,
\begin{align}
\mathrm{i}
\Gamma
\partial^{\,}_{\lambda}
|\Psi^{\,}_{\lambda}\rangle
=
\left(
H^{\,}_{0}+\lambda V
\right)
|\Psi^{\,}_{\lambda}\rangle,
\quad
|\Psi^{\,}_{0}\rangle=|\chi^{\,}_{0}\rangle,
\label{eq: reduced single body SE}
\end{align}
given 
$
H^{\,}_{0}
|\chi^{\,}_{n}\rangle
=
\varepsilon^{\,}_{n}|\chi^{\,}_{n}\rangle.
$
The following perturbative expansion in $\lambda$ (i.e., reduced time) is different from the usual time-dependent perturbation theory in which the expansion parameter
is time-independent.
Hence, we provide some details for our perturbative approach.
Generically, we can decompose $|\Psi^{\,}_{\lambda}\rangle$ as
\begin{align}
|\Psi^{\,}_{\lambda}\rangle=
\sum^{\,}_{n}
C^{\,}_{n}(\lambda)
\exp\left(-\mathrm{i}\lambda\varepsilon^{\,}_{n}/\Gamma\right)
|\chi^{\,}_{n}\rangle,
\label{eq: lambda expansion of psi}
\end{align}
with $\lambda$-dependent coefficients $C^{\,}_{n}(\lambda)$ from which a factor $\exp\left(-\mathrm{i}\lambda\varepsilon^{\,}_{n}/\Gamma\right)$ has
been extracted for later convenience. 
Since $|\Psi^{\,}_{0}\rangle=|\chi^{\,}_{0}\rangle,$
we have $C^{\,}_{n}(0)=\delta^{\,}_{n0}.$ 
Bring the decomposition (\ref{eq: lambda expansion of psi}) into Eq.\ (\ref{eq: reduced single body SE}) 
yields a first-order differential equation for $C^{\,}_{n}(\lambda),$
\begin{align}
\partial^{\,}_{\lambda}C^{\,}_{m}(\lambda)
=
\sum^{\,}_{n}C^{\,}_{n}(\lambda)
\exp\left(\mathrm{i}\lambda\omega^{\,}_{mn}/\Gamma\right)
\,\lambda 
\frac{V^{\,}_{mn}}{\mathrm{i}\Gamma},
\label{dynamical equation for coefficient}
\end{align}
where $\omega^{\,}_{mn}\:=\varepsilon^{\,}_{m}-\varepsilon^{\,}_{n}.$
Now, as we are interested in small $\lambda$ region, 
we may expand $C^{\,}_{n}(\lambda)$ in power series of $\lambda$,
namely,
$
C^{\,}_{n}(\lambda)
=\sum^{\infty}_{j=0}\lambda^{j}C^{(j)}_{n}
=\delta^{\,}_{n0}+\sum^{\infty}_{j=1}\lambda^{j}C^{(j)}_{n}.
$
We shall also expand the $\exp\left(\mathrm{i}\lambda\omega^{\,}_{mn}/\Gamma\right)$ factor in powers of $\lambda.$ 
The differential equation (\ref{dynamical equation for coefficient}) then reads
\begin{align}
&\sum^{\infty}_{j=1}
j\lambda^{j-1}C^{(j)}_{m}
=
\sum^{\,}_{n}
\left(
\sum^{\infty}_{j=0}\lambda^{j}C^{(j)}_{n}
\right)
\left(
\sum^{\infty}_{\ell=0}
\frac{1}{\ell!}
\left(
\mathrm{i}\lambda\omega^{\,}_{mn}/\Gamma
\right)^{\ell}
\right)
\lambda 
\frac{V^{\,}_{mn}}{\mathrm{i}\Gamma}.
\label{eq: time dependent expansion first try}
\end{align}

We now match terms for each order in $\lambda$.
One finds that $C^{(1)}_{m}
=
0$
and, generically,
the term in the $k$-th order of $\lambda$ with $k\geq1$
reads,
\begin{align}
\lambda^{k}:
\quad
C^{(k+1)}_{m}
=
\sum^{\,}_{n}
\sum^{k-1}_{\ell=0}
C^{(k-\ell-1)}_{n}
\frac{1}{\ell!}
\left(
\mathrm{i}\omega^{\,}_{mn}/\Gamma
\right)^{\ell}
\frac{V^{\,}_{mn}}{(k+1)\mathrm{i}\Gamma}.
\nonumber
\end{align}
The first few leading order contributions are
\begin{subequations}
\label{eq: perturbative coeffifient}
\begin{align}
&
C^{(2)}_{m}
=
\frac{V^{\,}_{m0}}{2\mathrm{i}\Gamma},
\quad
C^{(3)}_{m}
=
\frac{\omega^{\,}_{m0}V^{\,}_{m0}}{3\Gamma^2},
\quad
C^{(4)}_{m}
=
-
\sum^{\,}_{n}
\frac{V^{\,}_{n0}V^{\,}_{mn}}{8\Gamma^2}
-
\frac{\omega^{2}_{m0}V^{\,}_{m0}}{8\mathrm{i}\Gamma^3},
\\
&
C^{(5)}_{m}
=
\sum^{\,}_{n}
\left(
\frac{\omega^{\,}_{n0}}{3}
+
\frac{\omega^{\,}_{mn}}{2}
\right)
\frac{V^{\,}_{n0}V^{\,}_{mn}}{5\mathrm{i}\Gamma^3}
-
\frac{\omega^{3}_{m0}V^{\,}_{m0}}{30\Gamma^4}.
\end{align}
\end{subequations}

Upon substituting Eq.\ (\ref{eq: perturbative coeffifient}) into Eq.\ (\ref{eq: lambda expansion of psi}),
expanding terms up to order $\lambda^{5}$,
and separating terms into $n=0$ and $n\neq0$ yields
\begin{align}
|\Psi^{\,}_{\lambda}\rangle
=&\,
\Bigg[
1
+
\lambda
\frac{\varepsilon^{\,}_{0}}{\mathrm{i}\Gamma}
-
\lambda^2 
\left(
\frac{\varepsilon^{2}_{0}}{2\Gamma^2}
-
\frac{V^{\,}_{00}}{2\mathrm{i}\Gamma}
\right)
-
\lambda^3 
\left(
\frac{\varepsilon^{3}_{0}}{6\mathrm{i}\Gamma^3}
+
\frac{V^{\,}_{00}\varepsilon^{\,}_{0}}{2\Gamma^2}
\right)
+
\lambda^4 
\left(
\frac{\varepsilon^{4}_{0}}{24\Gamma^4}
-
\sum^{\,}_{m}
\frac{|V^{\,}_{m0}|^2}{8\Gamma^2}
-
\frac{V^{\,}_{00}\varepsilon^{2}_{0}}{4\mathrm{i}\Gamma^3}
\right)
\nonumber\\
&\quad
+
\lambda^5 
\left(
\frac{\varepsilon^{5}_{0}}{120\mathrm{i}\Gamma^5}
-
\sum^{\,}_{n}
\frac{\omega^{\,}_{n0}|V^{\,}_{n0}|^{2}}{30\mathrm{i}\Gamma^3}
-
\sum^{\,}_{n}
\frac{|V^{\,}_{n0}|^{2}\varepsilon^{\,}_{0}}{8\mathrm{i}\Gamma^3}
+
\frac{V^{\,}_{00}\varepsilon^{3}_{0}}{12\Gamma^4}
\right)
\Bigg]
|\chi^{\,}_{0}\rangle
\nonumber\\
&
+
\sum^{\,}_{n\neq0}
\Bigg[
\lambda^2 
\frac{V^{\,}_{n0}}{2\mathrm{i}\Gamma}
+
\lambda^3 
\left(
\frac{\omega^{\,}_{n0}V^{\,}_{n0}}{3\Gamma^2}
-
\frac{V^{\,}_{n0}\varepsilon^{\,}_{n}}{2\Gamma^2}
\right)
\nonumber\\
&\quad
+
\lambda^4 
\left(
-
\sum^{\,}_{m}
\frac{V^{\,}_{m0}V^{\,}_{nm}}{8\Gamma^2}
-
\frac{\omega^{2}_{n0}V^{\,}_{n0}}{8\mathrm{i}\Gamma^3}
+
\frac{\omega^{\,}_{n0}V^{\,}_{n0}\varepsilon^{\,}_{n}}{3\mathrm{i}\Gamma^3}
-
\frac{V^{\,}_{n0}\varepsilon^{2}_{n}}{4\mathrm{i}\Gamma^3}
\right)
+
\cdots
\Bigg]
|\chi^{\,}_{n}\rangle.
\end{align}

Thus, we obtain the following inner products,
\begin{subequations}
\label{eq: conclusion for time evolved perturbation}
\begin{align}
\langle\chi^{\,}_{0}|\Psi^{\,}_{\lambda}\rangle
=
&\,
1
+
\lambda
\frac{\varepsilon^{\,}_{0}}{\mathrm{i}\Gamma}
-
\lambda^2 
\left(
\frac{\varepsilon^{2}_{0}}{2\Gamma^2}
-
\frac{V^{\,}_{00}}{2\mathrm{i}\Gamma}
\right)
-
\lambda^3 
\left(
\frac{\varepsilon^{3}_{0}}{6\mathrm{i}\Gamma^3}
+
\frac{V^{\,}_{00}\varepsilon^{\,}_{0}}{2\Gamma^2}
\right)
\nonumber\\
&\,
+
\lambda^4 
\left(
\frac{\varepsilon^{4}_{0}}{24\Gamma^4}
-
\sum^{\,}_{m}
\frac{|V^{\,}_{m0}|^2}{8\Gamma^2}
-
\frac{V^{\,}_{00}\varepsilon^{2}_{0}}{4\mathrm{i}\Gamma^3}
\right)
\nonumber\\
&\,
+
\lambda^5 
\left(
\frac{\varepsilon^{5}_{0}}{120\mathrm{i}\Gamma^5}
-
\sum^{\,}_{m}
\frac{\omega^{\,}_{m0}|V^{\,}_{m0}|^{2}}{30\mathrm{i}\Gamma^3}
-
\sum^{\,}_{m}
\frac{|V^{\,}_{m0}|^{2}\varepsilon^{\,}_{0}}{8\mathrm{i}\Gamma^3}
+
\frac{V^{\,}_{00}\varepsilon^{3}_{0}}{12\Gamma^4}
\right)
+\cdots,
\label{eq: conclusion for time evolved perturbation a}
\\
\langle\chi^{\,}_{n\neq0}|\Psi^{\,}_{\lambda}\rangle
=&\,
\lambda^2 
\frac{V^{\,}_{n0}}{2\mathrm{i}\Gamma}
+
\lambda^3 
\left(
\frac{\omega^{\,}_{n0}V^{\,}_{n0}}{3\Gamma^2}
-
\frac{V^{\,}_{n0}\varepsilon^{\,}_{n}}{2\Gamma^2}
\right)
\nonumber\\
&\,
+
\lambda^4 
\left(
-
\sum^{\,}_{m}
\frac{V^{\,}_{m0}V^{\,}_{nm}}{8\Gamma^2}
-
\frac{\omega^{2}_{n0}V^{\,}_{n0}}{8\mathrm{i}\Gamma^3}
+
\frac{\omega^{\,}_{n0}V^{\,}_{n0}\varepsilon^{\,}_{n}}{3\mathrm{i}\Gamma^3}
-
\frac{V^{\,}_{n0}\varepsilon^{2}_{n}}{4\mathrm{i}\Gamma^3}
\right)
+
\cdots.
\label{eq: conclusion for time evolved perturbation b}
\end{align}
\end{subequations}

\subsection{Various overlaps in perturbative expansion}

We are ready to compute various overlaps using Eqs.\ (\ref{eq: conclusion for instantaneous perturbation}) and (\ref{eq: conclusion for time evolved perturbation}).
First, the ground state overlap $\mathcal{C}(\lambda)$ follows from Eq.\ (\ref{eq: conclusion for instantaneous perturbation a}),
\begin{align}
\mathcal{C}(\lambda)
=&\,
1
-
\lambda^{2}
\sum^{\,}_{n\neq 0}
\frac{|V^{\,}_{n0}|^2}{(\varepsilon^{\,}_{0}-\varepsilon^{\,}_{n})^2}
\nonumber\\
&\,
+
2
\lambda^{3}
\Bigg(
\sum^{\,}_{n\neq 0}
\frac{V^{\,}_{00}|V^{\,}_{n0}|^2}{(\varepsilon^{\,}_{0}-\varepsilon^{\,}_{n})^{3}}
-
\sum^{\,}_{n\neq 0}
\frac{1}{(\varepsilon^{\,}_{0}-\varepsilon^{\,}_{n})^2}
\sum^{\,}_{m\neq 0}
\frac{\Re(V^{*}_{nm}V^{*}_{m0}V^{\,}_{n0})}{\varepsilon^{\,}_{0}-\varepsilon^{\,}_{m}}
\Bigg)
+
\cdots.
\label{eq: perturbative GOC}
\end{align}

Second, the adiabatic fidelity $\mathcal{F}(\lambda)$ follows from Eqs.\ (\ref{eq: conclusion for instantaneous perturbation}) 
and (\ref{eq: conclusion for time evolved perturbation}),
\begin{align}
\mathcal{F}(\lambda)
=&\,|\langle\Phi^{\,}_{\lambda}|\Psi^{\,}_{\lambda}\rangle|^2
=
\Big|
\langle\Phi^{\,}_{\lambda}|\Phi^{\,}_{0}\rangle
\langle\Phi^{\,}_{0}|\Psi^{\,}_{\lambda}\rangle
+
\sum^{\,}_{n\neq 0}
\underbrace{
\langle\Phi^{\,}_{\lambda}|\chi^{\,}_{n}\rangle
\langle\chi^{\,}_{n}|\Psi^{\,}_{\lambda}\rangle
}^{\,}_{=\mathcal{O}(\lambda^3)}
\Big|^2
\nonumber\\
=&\,
1
-
\lambda^{2}
\sum^{\,}_{n\neq 0}
\frac{|V^{\,}_{n0}|^2}{(\varepsilon^{\,}_{0}-\varepsilon^{\,}_{n})^2}
\nonumber\\
&\,
+
2
\lambda^{3}
\Bigg(
\sum^{\,}_{n\neq 0}
\frac{V^{\,}_{00}|V^{\,}_{n0}|^2}{(\varepsilon^{\,}_{0}-\varepsilon^{\,}_{n})^{3}}
-
\sum^{\,}_{n\neq 0}
\frac{1}{(\varepsilon^{\,}_{0}-\varepsilon^{\,}_{n})^2}
\sum^{\,}_{m\neq 0}
\frac{\Re(V^{*}_{nm}V^{*}_{m0}V^{\,}_{n0})}{\varepsilon^{\,}_{0}-\varepsilon^{\,}_{m}}  
-
\frac{V^{\,}_{00}\varepsilon^{\,}_{0}}{2\Gamma^2}
\Bigg)
+\cdots,
\label{eq: perturbative AF}
\end{align}
which is identical to $\mathcal{C}(\lambda)$ (\ref{eq: perturbative GOC}) for up to order $\lambda^2$.
Their difference, $\mathcal{F}(\lambda)-\mathcal{C}(\lambda)$, reads
$
\mathcal{F}(\lambda)-\mathcal{C}(\lambda)=
-
\lambda^{3}
\frac{V^{\,}_{00}\varepsilon^{\,}_{0}}{\Gamma^2}+\cdots.
$

Third, $\cos^2\theta(\lambda)$ follows from Eq.\ (\ref{eq: conclusion for time evolved perturbation a}),
\begin{align}
\cos^2\theta(\lambda)
=
|\langle\Phi^{\,}_{0}|\Psi^{\,}_{\lambda}\rangle|^2
=
1
-
\frac{\lambda^4}{4}
\sum^{\,}_{n\neq0}
\frac{|V^{\,}_{n0}|^2}{\Gamma^2}
+
\lambda^{5}
\frac{V^{\,}_{00}\varepsilon^{3}_{0}}{\Gamma^4}
+
\cdots.
\end{align}
It then follows that $\sin^2\theta(\lambda)$ and $\sin\theta(\lambda)$ read
\begin{subequations}
\begin{align}
&\sin^2\theta(\lambda)
=
1-\cos^2\theta(\lambda)
=
\frac{\lambda^4}{4}
\sum^{\,}_{n\neq0}
\frac{|V^{\,}_{n0}|^2}{\Gamma^2}
-
\lambda^{5}
\frac{V^{\,}_{00}\varepsilon^{3}_{0}}{\Gamma^4}
+\cdots,
\label{eq: perturbative sinth square}
\\
&\sin\theta(\lambda)
=
\frac{\lambda^2}{2}
\left(
\sum^{\,}_{n\neq0}
\frac{|V^{\,}_{n0}|^2}{\Gamma^2}
\right)^{1/2}
-
\lambda^{3}
\left(
\sum^{\,}_{n\neq0}
\frac{|V^{\,}_{n0}|^2}{\Gamma^2}
\right)^{-1/2}
\frac{V^{\,}_{00}\varepsilon^{3}_{0}}{\Gamma^4}
+\cdots.
\label{eq: perturbative sinth}
\end{align}
\end{subequations}

Fourth, the overlap $\sqrt{\mathcal{D}^{\,}_{\mathrm{un}}(\lambda)}$ follows from Eqs.\ (\ref{eq: conclusion for instantaneous perturbation b}) 
and (\ref{eq: conclusion for time evolved perturbation}),
\begin{subequations}
\label{eq: perturbative Dun}
\begin{align}
\mathcal{D}^{\,}_{\mathrm{un}}(\lambda)
=
|\langle\Phi^{\,}_{\lambda}|\left(\mathbb{I}-P\right)|\Psi^{\,}_{\lambda}\rangle|^2
=
\left|
\sum^{\,}_{n\neq0}
\langle\Psi^{\,}_{\lambda}|\chi^{\,}_{n}\rangle\langle\chi^{\,}_{n}|\Phi^{\,}_{\lambda}\rangle
\right|^2
\=:
\lambda^{6}a^{\,}_{6}
+
\lambda^{7}a^{\,}_{7}
+\cdots,
\end{align}
where
\begin{align}
&
a^{\,}_{6}\:=
\frac{1}{4\Gamma^2}
\left(
\sum^{\,}_{n\neq0}
\frac{|V^{\,}_{n0}|^2}{\varepsilon^{\,}_{0}-\varepsilon^{\,}_{n}}
\right)^{2},
\\
&
a^{\,}_{7}\:=
\frac{1}{2\Gamma^2}
\left(
\sum^{\,}_{n\neq0}
\frac{|V^{\,}_{n0}|^2}{\varepsilon^{\,}_{0}-\varepsilon^{\,}_{n}}
\right)
\sum^{\,}_{n\neq0}
\frac{1}{\varepsilon^{\,}_{0}-\varepsilon^{\,}_{n}}
\left(
\sum^{\,}_{m\neq 0}
\frac{\Re\left(V^{*}_{nm}V^{*}_{m0}V^{\,}_{n0}\right)}{\varepsilon^{\,}_{0}-\varepsilon^{\,}_{m}}
-
\frac{V^{\,}_{00}|V^{\,}_{n0}|^2}{\varepsilon^{\,}_{0}-\varepsilon^{\,}_{n}}
\right).
\end{align}
\end{subequations}

Finally, we calculate $\sin^2\theta(\lambda)\left(1-\mathcal{C}(\lambda)\right)$ using Eqs.\ (\ref{eq: perturbative GOC}) and (\ref{eq: perturbative sinth square}):
\begin{subequations}
\label{eq: sqrt C sinth perturbative}
\begin{align}
\sin^2\theta(\lambda)\left(1-\mathcal{C}(\lambda)\right)
\=:
\lambda^{6}b^{\,}_{6}+\lambda^{7}b^{\,}_{7}
+\cdots,
\end{align}
where
\begin{align}
b^{\,}_{6}\:=&\,
\frac{1}{4}
\left(
\sum^{\,}_{n\neq0}
\frac{|V^{\,}_{n0}|^2}{\Gamma^2}
\right)
\left(
\sum^{\,}_{n\neq 0}
\frac{|V^{\,}_{n0}|^2}{(\varepsilon^{\,}_{0}-\varepsilon^{\,}_{n})^2}
\right),
\\
b^{\,}_{7}\:=\,&
-
\frac{V^{\,}_{00}\varepsilon^{3}_{0}}{\Gamma^4}
\sum^{\,}_{n\neq 0}
\frac{|V^{\,}_{n0}|^2}{(\varepsilon^{\,}_{0}-\varepsilon^{\,}_{n})^2}
\nonumber\\
&\,
-
\frac{1}{2}
\left(
\sum^{\,}_{n\neq0}
\frac{|V^{\,}_{n0}|^2}{\Gamma^2}
\right)
\Bigg(
\sum^{\,}_{n\neq 0}
\frac{V^{\,}_{00}|V^{\,}_{n0}|^2}{(\varepsilon^{\,}_{0}-\varepsilon^{\,}_{n})^{3}}
-
\sum^{\,}_{n\neq 0}
\frac{1}{(\varepsilon^{\,}_{0}-\varepsilon^{\,}_{n})^2}
\sum^{\,}_{m\neq 0}
\frac{\Re(V^{*}_{nm}V^{*}_{m0}V^{\,}_{n0})}{\varepsilon^{\,}_{0}-\varepsilon^{\,}_{m}}
\Bigg).
\end{align}
\end{subequations}
Combing Eq.\ (\ref{eq: sqrt C sinth perturbative}) with Eq.\ (\ref{eq: perturbative Dun}) and Eq.\ (\ref{eq: relation between two D overlap})
yields
\begin{align}
\mathcal{D}(\lambda)
=&
\frac{
\left(
\sum^{\,}_{n\neq0}
\frac{|V^{\,}_{n0}|^2}{\varepsilon^{\,}_{0}-\varepsilon^{\,}_{n}}
\right)^{2}
}
{
\left(
\sum^{\,}_{n\neq0}
|V^{\,}_{n0}|^2
\sum^{\,}_{m\neq 0}
\frac{|V^{\,}_{m0}|^2}{(\varepsilon^{\,}_{0}-\varepsilon^{\,}_{m})^2}
\right)
}
\nonumber\\
&\,
+
16\lambda
\left(
\sum^{\,}_{n\neq0}
\frac{|V^{\,}_{n0}|^2}{\Gamma^2}
\sum^{\,}_{m\neq 0}
\frac{|V^{\,}_{m0}|^2}{(\varepsilon^{\,}_{0}-\varepsilon^{\,}_{m})^2}
\right)^{-2}
\left(
a^{\,}_{7}b^{\,}_{6}-a^{\,}_{6}b^{\,}_{7}
\right)
+\cdots.
\label{eq: perturbative sqrt D}
\end{align}

\section{Non-interacting Hamiltonians}
\label{App2: Non-interacting Hamiltonians}


We consider non-interacting systems
whose Hamiltonian can be written as $N$-commuting pieces in momentum space,
i.e.,
$
H^{\,}_{\lambda}=\bigoplus^{N}_{k=1}\mathcal{H}^{\,}_{\lambda}(k).
$
Correspondingly, both the instantaneous ground state $|\Phi^{\,}_{\lambda}\rangle$
and the time-evolved state $|\Psi^{\,}_{\lambda}\rangle$
can be written as a tensor product form,
\begin{align}
|\Phi^{\,}_{\lambda}\rangle
=
\bigotimes^{N}_{k=1}
|\phi^{\,}_{\lambda}(k)\rangle,
\qquad
\text{and}
\qquad
|\Psi^{\,}_{\lambda}\rangle
=
\bigotimes^{N}_{k=1}
|\psi^{\,}_{\lambda}(k)\rangle,
\end{align}
where $|\phi^{\,}_{\lambda}(k)\rangle$ is the instantaneous ground state of $\mathcal{H}^{\,}_{\lambda}(k)$,
whereas for each $k$,
$|\psi^{\,}_{\lambda}(k)\rangle$ solves
\begin{align}
\mathrm{i}\Gamma\partial^{\,}_{\lambda}|\psi^{\,}_{\lambda}(k)\rangle
=
\mathcal{H}^{\,}_{\lambda}(k)
|\psi^{\,}_{\lambda}(k)\rangle,
\quad
|\psi^{\,}_{0}(k)\rangle
=
|\phi^{\,}_{0}(k)\rangle.
\end{align}
It then follows that 
the overlaps of various many-body wavefunctions can be written as products of overlaps of single-body wavefunctions
\begin{align}
&
\langle\Psi^{\,}_{\lambda}|\Phi^{\,}_{\lambda}\rangle
=
\prod^{\,}_{k}\langle\psi^{\,}_{\lambda}(k)|\phi^{\,}_{\lambda}(k)\rangle,
\qquad
\langle\Phi^{\,}_{0}|\Phi^{\,}_{\lambda}\rangle
=
\prod^{\,}_{k}\langle\phi^{\,}_{0}(k)|\phi^{\,}_{\lambda}(k)\rangle.
\label{eq: various overlap}
\end{align}

Define the single-body projector
$
p^{\,}_{k}\:=|\phi^{\,}_{0}(k)\rangle\langle\phi^{\,}_{0}(k)|
$
and its complementary projector $q^{\,}_{k}\:=\mathbb{I}^{\,}_{k}-p^{\,}_{k}$,
and make use of Eqs.\ (\ref{eq: various overlap}), 
we can express $\sqrt{\mathcal{D}^{\,}_{\mathrm{un}}(\lambda)}$ (\ref{eq: define un normalized D overlap}) as follows
\begin{align}
&\sqrt{\mathcal{D}^{\,}_{\mathrm{un}}(\lambda)}=
\left|
\langle\Psi^{\,}_{\lambda}|P|\Phi^{\,}_{\lambda}\rangle
\right|
\left|
\prod^{\,}_{k}
\left(
1+
A^{\,}_{k}
\right)
-1
\right|,
\quad\text{where}\quad
A^{\,}_{k}\:=
\frac{
\langle\psi^{\,}_{\lambda}(k)|q^{\,}_{k}|\phi^{\,}_{\lambda}(k)\rangle
}{\langle\psi^{\,}_{\lambda}(k)|p^{\,}_{k}|\phi^{\,}_{\lambda}(k)\rangle}.
\label{eq: sDun vis}
\end{align}
To make further progress, a crucial observation for $A^{\,}_{k}$ is that, for each $k,$
the following condition holds
\begin{align}
|\langle\psi^{\,}_{\lambda}(k)|q^{\,}_{k}|\phi^{\,}_{\lambda}(k)\rangle|
\ll
|\langle\psi^{\,}_{\lambda}(k)|p^{\,}_{k}|\phi^{\,}_{\lambda}(k)\rangle|.
\label{eq: smallness condition}
\end{align}
This fact can be verified directly by considering a perturbative expansion in $\lambda$ similar to what has been done in App.\ \ref{App1: Perturbative  expansion}.
If so, the following approximation formula,
\begin{align}
\prod^{\,}_{k}(1+A^{\,}_{k})
\simeq
1+\sum^{\,}_{k}A^{\,}_{k}
\qquad
\text{for all } |A^{\,}_{k}|\ll1,
\label{eq: approximation formula for Ak}
\end{align}
can be applied to Eq.\ (\ref{eq: sDun vis}).
Upon using Eqs.\ (\ref{eq: properties by constructions}) and (\ref{eq: approximation formula for Ak}), Eq.\ (\ref{eq: sDun vis}) reads 
\begin{align}
\sqrt{\mathcal{D}^{\,}_{\mathrm{un}}(\lambda)}
&
\simeq
\cos\theta(\lambda)
\sqrt{\mathcal{C}(\lambda)}
\left|
\sum^{\,}_{k}
A^{\,}_{k}
\right|.
\label{eq: sDun vis general}
\end{align}

\subsection*{Driven Rice-Mele model}

We now apply the formalism developed above to the Rice-Mele model (\ref{eq: driven RM model}).
Upon performing a Fourier transform,
the Rice-Mele Hamiltonian (\ref{eq: driven RM model}) can be written as a sum of $N$ commuting terms
\begin{subequations}
\begin{align}
&
H^{\,}_{\mathrm{RM}}
=
\sum^{\,}_{k}
\begin{pmatrix}
a^{\dag}_{k}
&
b^{\dag}_{k}
\end{pmatrix}
\mathcal{H}^{\,}_{\lambda}(k)
\begin{pmatrix}
a^{\,}_{k}
\\
b^{\,}_{k}
\end{pmatrix},
\end{align}
where 
$
\mathcal{H}^{\,}_{\lambda}(k)
=
\bm{d}^{\,}_{\lambda}(k)\cdot\bm{\sigma}
$
and
\begin{align}
\bm{d}^{\,}_{\lambda}(k)\:=
\begin{pmatrix}
-(J+U)-(J-U)\cos k
\\
(J-U)\sin k
\\
\mu(\lambda)
\end{pmatrix}
\label{eq: define d vector app}
\end{align}
\end{subequations}
with $\bm{\sigma}$ are the Pauli matrices. 

We shall specialize to the case in which $J=U=$ constant and $\mu(\lambda)=\lambda.$
It then follows that the $\bm{d}$ vector (\ref{eq: define d vector app}) has no momentum dependence and
each single-body Hamiltonian $\mathcal{H}^{\,}_{\lambda}(k)$ (\ref{eq: define d vector app}) is simply the Landau-Zener model.
For this case, the $\left|\sum^{\,}_{k}A^{\,}_{k}\right|$ term in Eq.\ (\ref{eq: sDun vis general}) simplifies
\begin{align}
\left|
\sum^{\,}_{k}A^{\,}_{k}
\right|
&=
N
\frac{
\sqrt{1-\left|\langle\psi^{\,}_{\lambda}|\phi^{\,}_{0}\rangle\right|^2}
\sqrt{1-\left|\langle\phi^{\,}_{\lambda}|\phi^{\,}_{0}\rangle\right|^2}
}{
\left|\langle\psi^{\,}_{\lambda}|\phi^{\,}_{0}\rangle\right|
\left|\langle\phi^{\,}_{\lambda}|\phi^{\,}_{0}\rangle\right|
},
\label{eq: sDun vis two-level and k independent app}
\end{align}
where each overlap of single-body states can be obtained easily
\begin{align}
&
|\langle\phi^{\,}_{\lambda}|\phi^{\,}_{0}\rangle|
=
|\langle\Phi^{\,}_{\lambda}|\Phi^{\,}_{0}\rangle|^{\frac{1}{N}}
=
\left(\sqrt{\mathcal{C}(\lambda)}\right)^{\frac{1}{N}},
\qquad
|\langle\psi^{\,}_{\lambda}|\phi^{\,}_{0}\rangle|
=
|\langle\Psi^{\,}_{\lambda}|\Phi^{\,}_{0}\rangle|^{\frac{1}{N}}
=
\left(
\cos\theta(\lambda)
\right)^{\frac{1}{N}}.
\end{align}
Using these results,
Eq.\ (\ref{eq: sDun vis two-level and k independent app}) can be expressed in terms of $\theta(\lambda)$ and $\mathcal{C}(\lambda)$ as
\begin{align}
\left|
\sum^{\,}_{k}
A^{\,}_{k}
\right|
&=
N
\frac{
\sqrt{1-\left(
\cos^2\theta(\lambda)
\right)^{\frac{1}{N}}}
\sqrt{1-\left(\mathcal{C}(\lambda)\right)^{\frac{1}{N}}}
}{
\left(
\cos\theta(\lambda)
\right)^{\frac{1}{N}}
\left(\sqrt{\mathcal{C}(\lambda)}\right)^{\frac{1}{N}}
}
\quad\geq\quad
\sqrt{N}
\frac{
\sin\theta(\lambda)
\sqrt{1-\left(\mathcal{C}(\lambda)\right)^{\frac{1}{N}}}
}{
\left(
\cos\theta(\lambda)
\right)^{\frac{1}{N}}
\left(\sqrt{\mathcal{C}(\lambda)}\right)^{\frac{1}{N}}
},
\label{eq: sDun vis two-level and k independent final app}
\end{align}
where we have used the inequality $(1-x)^{n}\leq(1-nx)$ for $0\leq x\leq1$ and $0<n<1.$
It then follows that $\sqrt{\mathcal{D}^{\,}_{\mathrm{un}}(\lambda)}$ (\ref{eq: sDun vis general}) reads
\begin{subequations}
\label{eq: sDun vis final two level and k ind app}
\begin{align}
&\sqrt{\mathcal{D}^{\,}_{\mathrm{un}}(\lambda)}
\geq
\left(
\cos\theta(\lambda)
\right)^{1-\frac{1}{N}}
\left(
\sqrt{\mathcal{C}(\lambda)}
\right)^{1-\frac{1}{N}}
\sin\theta(\lambda)
\alpha(\lambda),
\label{eq: sDun vis final two level and k ind a app}
\\
&
\alpha(\lambda)\:=
\sqrt{N}\sqrt{1-\left(\mathcal{C}(\lambda)\right)^{\frac{1}{N}}},
\label{eq: sDun vis final two level and k ind b app}
\end{align}
\end{subequations}
where the equality in Eq.\ (\ref{eq: sDun vis final two level and k ind a app}) holds if
$\theta(\lambda)$ is small.
Note that the exponent $1-\frac{1}{N}$ in Eq.\ (\ref{eq: sDun vis final two level and k ind a app}) may be approximated as $1$ if $N$ is large.

\section{Derivation of Eq.\ (\ref{eq: scaling down of Gamma general form})}
\label{App3: Detail calculation for implication on adiabaticity breakdown}

Combing triangle inequality (\ref{eq: a more useful form for  triangle inequality})
and the inequality
$
1-\sqrt{\mathcal{F}(\lambda)}\leq\epsilon
$
with 
$
\epsilon\in[0,1]
$
from quantum adiabatic theorem,
we obtain 
\begin{align}
\left(1-\epsilon-\sqrt{\mathcal{C}(\lambda)}\right)^{2}
&\leq
\mathcal{D}^{\,}_{\mathrm{un}}(\lambda)
\nonumber\\
&\overset{(\ref{eq: relation between two D overlap})}{=}
\sin\theta(\lambda)
\sqrt{
1-\mathcal{C}(\lambda)
}
\sqrt{\mathcal{D}(\lambda)}
\nonumber\\
&\overset{(\ref{eq: sD postulate form})}{=}
\sin\theta(\lambda)
\sqrt{
1-\mathcal{C}(\lambda)
}
\sqrt{\mathcal{C}(\lambda)^{s}}
\nonumber\\
&\overset{(\ref{eq: bound from QSL})}{\leq}
\sin\widetilde{\mathcal{R}}(\lambda)
\sqrt{
1-\mathcal{C}(\lambda)
}
\sqrt{\mathcal{C}(\lambda)^{s}},
\end{align}
where 
$C(\lambda)=e^{-C^{\,}_{N}\lambda^{2}}$ as $N\to \infty.$
We shall take $\lambda=\lambda^{\,}_{*}=C^{-1/2}_{N}$ in the inequality above.
Since we are interested in the limit where $\delta V^{\,}_{N}/C^{\,}_{N}\to0$ as $N\to\infinity,$
we may approximate $\sin\widetilde{\mathcal{R}}(\lambda^{\,}_{*})$
by $\mathcal{R}(\lambda^{\,}_{*})=\delta V^{\,}_{N}/(2\Gamma C^{\,}_{N})$ from Eq.\ (\ref{eq: bound from QSL special form}),
\begin{align}
\left(1-\epsilon-\sqrt{\mathcal{C}(\lambda^{\,}_{*})}\right)^{2}
&\leq
\sin\widetilde{\mathcal{R}}(\lambda^{\,}_{*})
\sqrt{
1-\mathcal{C}(\lambda^{\,}_{*})
}
\sqrt{\mathcal{C}(\lambda^{\,}_{*})^{s^{\,}_{*}}}
\nonumber\\
&\lesssim
\mathcal{R}(\lambda^{\,}_{*})
\sqrt{
1-e^{-1}
}
e^{-s^{\,}_{*}/2},
\label{eq: scaling down of Gamma general form appendix}
\end{align}
where
$s^{\,}_{*}\:=s(\lambda^{\,}_{*})=-\ln\mathcal{D}(\lambda^{\,}_{*}).$
Equation (\ref{eq: scaling down of Gamma general form appendix}) implies
\begin{align}
&\Gamma
\leq
\frac{1}{2}
\frac{\delta V^{\,}_{N}}{C^{\,}_{N}}
\frac{1}{\left(1-\epsilon-e^{-1/2}\right)^{2}}
M(s^{\,}_{*}),
\end{align}
where
$
M(s^{\,}_{*})\:=
\sqrt{
1-e^{-1}
}
e^{-s^{\,}_{*}/2}.
$
\end{appendix}



\bibliography{references-adiabatic-closeness.bib}

\nolinenumbers

\end{document}